\documentclass[twocolumn,english,aps,prx,floatfix,amssymb,superscriptaddress,longbibliography]{revtex4}
\usepackage[latin9]{inputenc}
\usepackage{verbatim}
\usepackage{float}
\usepackage{amsmath}
\usepackage{amssymb}
\usepackage{graphicx}
\usepackage{color}
\usepackage{xcolor}
\usepackage{tikz}
\newcommand{\ket}[1]{\ensuremath{\left| #1 \right>}}

\usepackage{bbold}

\usepackage[bookmarks=false,linkcolor=blue,urlcolor=blue,colorlinks,citecolor=blue]{hyperref}
\makeatletter

\newcommand{\be}{\begin{equation}}
\newcommand{\ee}{\end{equation}}
\newcommand{\bea}{\begin{eqnarray}}
\newcommand{\eea}{\end{eqnarray}}

\begin{document}

\title{Boundary Quantum Knizhnik-Zamolodchikov Equations and Integrability of Quantum Field Theories with Time-Dependent Bulk and Boundary Coupling Strengths}
\author{Parameshwar R. Pasnoori}
\affiliation{Condensed Matter Theory Center, Department of Physics, University of Maryland, College Park, MD 20742, USA}
\affiliation{Mani L. Bhaumik Institute for Theoretical Physics, Department of Physics and Astronomy, \\ University of California, Los Angeles, CA 90095, USA}

\email{parmesh12@g.ucla.edu}
\begin{abstract}
The generalized Bethe ansatz framework formulated in [P. R. Pasnoori, Phys. Rev. B 112, L060409 (2025)]
provides a unified framework to find exact solutions to quantum many-body systems with time-dependent coupling strengths with periodic boundary conditions. In this work we extend this framework to the case of open boundary conditions where in addition to the time-dependent interactions in the bulk, the boundary conditions are explicitly time-dependent. We show that for integrable time-dependent bulk coupling strengths, the generalized Bethe ansatz framework provides the time-dependent boundary conditions compatible with integrability and reduces the time-dependent Schrodinger equation to a set of matrix difference equations called the boundary quantum Knizhnik-Zamolodchikov (bqKZ) equations. The solution to the bqKZ equations provides the explicit form of the exact wavefunction. We further show that the RG invariants of the corresponding static model identify with the dynamical invariants in the time-dependent model.
\end{abstract}
\maketitle

\section{Introduction}
Boundaries play a crucial role in identifying novel phases of matter \cite{BKT,Thouless,Haldane,Laughlin,Kane,Pollman,Keselman} and in some instances completely capture the dynamics of the bulk through the bulk boundary correspondence \cite{Yasuhiro,Halperin,NHbbcor,kellendonk}. In this regard integrability offers powerful mathematical techniques which helps one to probe the boundary phenomena by analyzing the bulk and vice versa \cite{XXZpaper,PAA1,PAA2,pasnoori2025duality,pasnoori2025interplay}. Bethe ansatz framework both in the coordinate \cite{Bethe1931,Hulthen} and algebraic incarnations \cite{SklyaninQISM} provides the exact solution to integrable systems with constant coupling strengths, using which one can extract physical quantities exactly even in the presence of strong interactions \cite{Andrei80,AndreiLowenstein79,Thacker,WiegmannGN,Wiegmann_1981,TsvelickWiegmann1983}. The Bethe ansatz was extended to include the case of open boundaries in \cite{Cherednik1,Sklyanin}, which has led to extensive works in identifying integrable boundary conditions \cite{DeVega,BATCHELOR,LIMASANTOS,INAMI,Malara_2006,Lima-Santos_2009}. One can make use of the methods developed in the seminal works mentioned above to probe novel non perturbative phenomena. This program has been fruitful in identifying novel boundary effects \cite{XXZpaper,PAA1,PAA2,PasnooriXXZPD, KondoXXX,XXZKondo}, boundary phase transitions \cite{XXXpaper,PAA2,SUSYpaper,parmeshkondo1,parmeshkondo2} and has also found applications in quantum information science \cite{circuitSPT,EVcircuitSM,EVtrotter}. 

All of the above mentioned phenomena pertains to systems with constant coupling strengths. When the coupling strengths are explicitly time-dependent, the standard Bethe ansatz is no longer applicable. To overcome this, a generalized Bethe ansatz framework was developed in \cite{PasnooriKondo}. In this framework one constructs an exact wavefunction which is the solution to the time-dependent Schrodinger equation. Just as in the regular Bethe ansatz, the wavefunction consists of different ordering of the particles in the configuration space. The amplitude corresponding to one particular ordering of the particles and another amplitude which differs by the ordering of any particular pair of particles are related through an S-matrix, which exchanges these particles. The algebra of the generators governing the bulk interactions fixes the structure of this S-matrix. Unlike the systems with constant coupling strengths, the S-matrix dependents on both the time and spatial coordinates of the corresponding particles. These S-matrices satisfy the Yang-Baxter equation which is a necessary condition for integrability. One amplitude can be related to any amplitude in the wavefunction through a `transport operator' which consists of the string of S-matrices corresponding to the sequence of the exchange of particles which connect the two amplitudes. Certainly, there exists multiple paths connecting any two amplitudes.  Integrability requires that the transport operators corresponding to any path connecting the two amplitudes should commute, which is ensured by the Yang-Baxter equation. An equivalent way to implement this constraint is to require that transporting one particle across the system and then transporting another particle should commute with the transportation of the particles through the system in the reverse order. This is nothing but the zero curvature condition. When periodic boundary conditions are applied, the transport operators which transport the respective particles throughout the system once take the form of quantum Knizhnik-Zamolodchikov (qKZ) equations \cite{rishetikhin1,Smirnov_1986,rishetikhin2}, which are matrix difference equations associated with the amplitudes in the many-body wavefunction. In contrast, in the case of regular Bethe ansatz framework, the transport operator which consists of string of S-matrices which are independent of time and the positions of the particles are related to the transfer matrix. When appropriate boundary conditions are applied, one obtains an eigenvalue problem which produces Bethe equations. n the case of time-dependent strengths, the zero curvature condition associated with the qKZ equations, which ensures the consistency of the wavefunction provides constraints on the time-dependent coupling strengths. For time-dependent strengths satisfying these constraints the system is integrable and the solutions to the qKZ equations \cite{Frenkel,rishetikhin1,rishetikhin2,Babujian_1997,Tarasov:1994bb,varchenko} provides the explicit form of the many-body wavefunction \cite{Pasnoorigrossneveu,pasnoori2025Kondo2}, using which one can extract exact dynamical phenomena \cite{pasnoorigrossneveu2}.

 Remarkably, the constraints imposed by integrability where shown to be exactly equivalent to the renormalization group equations of the corresponding static model when time `$t$' is identified with the logarithm of the cutoff `$\log\Lambda$' \cite{PasnooriRG}. This connection between time-dependent coupling strengths and the RG flow was identified prior to the development of the generalized Bethe ansatz in \cite{Benhoare}  by showing that the time-dependent theory admits a Lax connection thereby preserving integrability when the time-dependent coupling strengths satisfy the one-loop RG equations of the static theory. This connection was also found in a recent work \cite{komatsu2026} through the connection between integrability and gauge theories \cite{Costello1,Costello2,costello3}. It was recently found in \cite{pasnooriNHKtime} using the generalized Bethe ansatz that, in the case of non Hermitian systems, the class of integrability preserving time-dependent strengths is larger than the one dictated by the integrability-RG flow connection.

In this work we extend this framework to the case of open boundaries with time-dependent boundary conditions. We show that just as in the case of integrable systems with constant bulk interaction strengths, where the regular Bethe ansatz provides the associated boundary conditions compatible with integrability, here we show that the generalized Bethe ansatz imposes constraints on the time-dependent boundary conditions for given integrable time-dependent bulk interaction strengths. Just as in the case of periodic boundary conditions, the bulk S-matrices whose structure is fixed by the algebra of the generators of the bulk interactions satisfy the quantum Yang-Baxter equation. The time-dependent boundary conditions determine the structure of the boundary S-matrix which connects the two amplitudes which differ by the chirality of the particle due to the reflection at the corresponding boundary. Integrability requires that the bulk and the boundary S-matrices satisfy reflection equation, which fixes the time-dependent boundary conditions for an integrable time-dependent bulk interaction strength. Here we note that in the static problems the bulk S-matrices and the boundary S-matrix corresponding to one boundary satisfy the reflection equation whereas the bulk S-matrices and the boundary S-matrix corresponding to the other boundary satisfy a dual reflection equation. As we shall see, in the case of time-dependent interactions, the reflection equations satisfied by the bulk and the boundary S-matrices corresponding to both the boundaries take the same form. Just as in the case of periodic boundary conditions, any amplitude in the exact many-body wavefunction is related to one amplitude of our choosing through the action of bulk and the boundary S-matrices. To determine this amplitude, and hence the exact form of the many-body wavefunction, we construct the transport operator which moves a particle around the system once, which now consists of both the bulk and the boundary S-matrices. We show that  the chosen amplitude satisfies a set of matrix difference equations called the boundary quantum Knizhnik-Zamolodchikov equations \cite{cherednikbqkz,RishetikhinbqKZ}. The solution to these equations determines the exact form of the amplitude and hence also of the complete wavefunction. The bqKZ equations, which are the extension of the quantum Knizhnik-Zamolodchikov (qKZ) equations for systems with boundaries have been well studied in the literature: The development of the qKZ equations evolved from early work on conformal field theories and integrable systems with boundaries. The original classical KZ equations were discovered by Knizhnik and Zamolodchikov \cite{KnizhnikZamolodchikov} which are differential equations satisfied by the correlation functions in CFTs associated with affine Lie algebra. Quantum variant of the KZ equations first appeared in the work of Smirnov \cite{Smirnov_1986} as fundamental equations for form-factors in the sine-Gordon model and were subsequently derived from the representation theory of quantum affine algebras by Frenkel and Reshetikhin \cite{Frenkel}. Solutions using Jackson integrals and off-shell Bethe vectors were initially established for quantum affine $\text{sl}_{2}$ by Reshetikhin \cite{rishetikhin1,rishetikhin2} and generalized to $\text{sl}_{n}$ by Tarasov and Varchenko \cite{Tarasov:1993vs}. The solutions to the qKZ equations was formalized in the language of off-shell Bethe ansatz by Bubujian et.al \cite{Babujian_1997}. Parallel to this, research into integrable systems with reflecting boundary conditions pioneered by Cherednik's reflection equations \cite{Cherednik1} and further developed by Sklyanin \cite{Sklyanin} paved the way for boundary qKZ equations \cite{cherednikbqkz}. This framework culminated in the finding that correlation functions and matrix elements of vertex operators with respect to boundary states satisfy these boundary qKZ equations explicitly \cite{BAJNOK2006179} and it has also found applications in combinatorics \cite{DiFrancesco_2007}. Thus, from a purely mathematical perspective, our work provides a novel and highly non trivial application of the bqKZ equations.

To demonstrate the generalized Bethe ansatz method, we provide the detailed construction for the most simplest non trivial example, which is the chiral invariant $SU(2)$ Gross-Neveu model or simply the $SU(2)$ Gross-Nevu model with time dependent bulk interaction strengths and time dependent boundary conditions. As we shall see, the $SU(2)$ symmetry leads to the rational R-matrix, which is associated with the S-matrices corresponding to the two particle scattering processes. We then discuss the anisotropic version, which is the $U(1)$ Thirring model. The S-matrices of this model are associated with the XXZ R-matrix which is a q-deformation associated with the quantum affine algebra $\mathcal{U}_q(\widehat{\text{sl}_2})$.  We stress that the method developed in this work is general and as we shall elaborate later in the manuscript, just as in the case of periodic boundary conditions, the method can be straightforwardly applied to identify and solve time-dependent Hamiltonians in which the generators of the interactions realize different algebras.

The Hamiltonian of the $U(1)$ Thirring model with time dependent strengths is given by 
 $H_{\text{TH}}=\int_{0}^{L}\mathrm{d}x \mathcal{H}_{\rm TH } (x,t)$, where the Hamiltonian density $\mathcal{H}_{\text{TH}}(x,t)$ takes the following form
\begin{align}\nonumber
&\mathcal{H}_{\text{TH}}(x,t)\hspace{-1mm}= \hspace{-2.5mm}\sum_{a=\uparrow,\downarrow}\hspace{-1.5mm}\psi^{\dagger}_{L a}(x,t)i\partial_x \psi^{}_{L a} (x,t)\hspace{-0.5mm}-\hspace{-0.5mm}\psi^{\dagger}_{R a}(x,t)i\partial_x \psi^{}_{R a}(x,t) \\\nonumber&+2\vec{A}_{ab,cd}(t)\hspace{-3mm}\sum_{a,b,c,d=\uparrow,\downarrow}\hspace{-2mm}\psi^{\dagger}_{Ra}(x,t)\psi^{\dagger}_{Lc}(x,t)\psi_{Rb}(x,t)\psi_{Ld}(x,t),\\&\text{where,}\;\;\vec{A}_{ab,cd}(t)=g_{\perp}(t)\left(\sigma^x_{ab}\sigma^x_{cd}+\sigma^y_{ab}\sigma^y_{cd}\right)+g_{\parallel}(t)\sigma^z_{ab}\sigma^z_{cd}.
\label{Hamiltonian}
\end{align}

Here, the fields $\psi_{L(R) a}(x,t)$, $a=(\uparrow, \downarrow)$, 
describe left and right moving fermions carrying spin $1/2$. We set their velocity $v_F=1$. The two terms in the first line describe the right and left moving fermions respectively and the third term describes the spin exchange interaction between a left and a right moving fermion as they cross. Here the interaction strengths $g_{\parallel}(t)$  and $g_{\perp}(t)$ are dependent on time and uniform throughout space.

The model exhibits $U(1)$ symmetry which corresponds to the conservation of spin along the $z$ direction: $s^z=\int_0^Ldx s^z(x,t)$, where

\begin{align}
\label{Sz}
 s^z(x,t)\hspace{-1mm}= \hspace{-1.3mm}\sum_{ab} \hspace{-1mm}\left(\psi^{\dagger}_{La}(x,t) \sigma^{z}_{ab} \psi_{Lb}(x,t) \hspace{-0.5mm}+\hspace{-0.5mm} \psi^{\dagger}_{Ra}(x,t) \sigma^{z}_{ab} \psi_{Rb}(x,t)\right).
\end{align}

In the special case where the interaction strengths are equal $g_{\parallel}(t)=g_{\perp}(t)$, the $U(1)$ Thirring model reduces to $SU(2)$ Gross-Neveu model, where the bulk is $SU(2)$ invariant as it commutes with the total spin operator $\vec{s}$, where
$\vec s = (1/2) \int_{0}^{L} dx\; \vec s(x,t)$ with
\begin{align}
\label{Spin}
\vec s(x,t)\hspace{-1mm}=\hspace{-1.3mm} \sum_{ab}\hspace{-1mm} \left(\psi^{\dagger}_{La}(x,t) \vec \sigma_{ab} \psi_{Lb}(x,t)\hspace{-0.5mm} +\hspace{-0.5mm} \psi^{\dagger}_{Ra}(x,t) \vec \sigma_{ab} \psi_{Rb}(x,t)\right).
\end{align}
The Hamiltonian also exhibits a discrete spin flip symmetry
\be \tau: \psi_{R\uparrow}(x,t)\rightarrow\psi_{R\downarrow}(x,t),\;\;  \psi_{L\uparrow}(x,t)\rightarrow\psi_{L\downarrow}(x,t), \;\; \tau^2=1.\ee

We consider the system on a line segment with boundary conditions that are time-dependent
\begin{align}\nonumber\psi_{La}(0,t)=-\mathcal{B}^{L}_{ab}(t)\psi_{Rb}(0,t), \\ \psi_{Ra}(L,t)=-\mathcal{B}^{R}_{ab}(t)\psi_{Lb}(L,t). \label{OBC}
\end{align}
The most general bulk interaction strength $g(t)$ which is compatible with integrability can be found through the generalized Bethe ansatz method by applying periodic boundary conditions \cite{Pasnoorigrossneveu,PasnooriU(1)Thirring,pasnoori2026sgcircuit}
\begin{align}f(t)=\alpha t+\beta,\label{intrelf}
\end{align}
where $f(t)$ is related to the bulk coupling strengths through a non universal relation. In the case of $U(1)$ Thirring model, the two coupling strengths $g_{\parallel}(t)$ and $g_{\perp}(t)$ are related to the two parameters $f(t)$ and $u$ through the relations \cite{pasnoori2026sgcircuit,PasnooriU(1)Thirring}
\begin{align}f(t)=\text{arccoth}\left\{\left(\frac{\sin^2{\frac{g_{\parallel}(t)}{2}}/\sin(\frac{1}{2}(g_{\parallel}(t)-g_{\perp}(t)))}{\sin(\frac{1}{2}(g_{\parallel}(t)+g_{\perp}(t))}\right)^{1/2}\right\}\label{utrel1},\end{align}
whereas the parameter $u$ is constant
 \begin{align} \cos u= \frac{\cos(g_{\parallel}(t))}{\cos(g_{\perp}(t))}\label{utrel2}.\end{align}
The case of $SU(2)$ Gross-Neveu model corresponds to the limit $u\rightarrow 0$, where
\begin{align}f(t)=\cot(g(t)/2).\label{gnrelr}\end{align}
Just as in the case of any integrable quantum field theory with constant couplings solvable by the standard Bethe ansatz, the relation between the parameters associated with integrability such as the spectral parameter and crossing parameter and the bare interaction strength in the Hamiltonian depends on the regularization scheme used. For example, another regularization scheme provides the following relation for the $SU(2)$ Gross-Neveu model
\begin{align}f(t)=\frac{1}{2g(t/2)}\left(1-\frac{3g(t/2)^2}{4}\right).\label{gnrel}\end{align}
In the universal regime, which generally corresponds to small values of the coupling strengths, the expressions corresponding to different regularization schemes (\ref{gnrelr}) and (\ref{gnrel}) coincide. In this universal regime, the above relations (\ref{utrel1}), (\ref{utrel2}), (\ref{gnrelr}) give rise to a remarkable connection between the integrable time-dependent strengths and the RG trajectories of the corresponding static model \cite{PasnooriRG} mentioned above. In this time-dependent integrability RG flow connection, the parameters that are dependent on time, such as the parameter $f(t)$, necessarily follow the RG trajectories of the static model and determine the time-dependent energy scales exhibited by the system \cite{pasnoorigrossneveu2}. In addition, the parameters that are independent of time, such as the parameter $u$, correspond to RG invariants in the static model. 

In the case of open boundary conditions, for the integrable bulk strengths $g_{\parallel}(t)$ and $g_{\perp}(t)$, we need to find the most general time-dependent boundary conditions (\ref{OBC}) which are compatible with integrability. In this work we only consider the case of diagonal boundaries, which break the $\mathbb{Z}_2$ spin flip symmetry and thus the $SU(2)$ spin rotation symmetry, but preserve the $U(1)$ symmetry associated with the total spin along the $z$ direction.

Under the open boundary conditions, the chirality of the particle changes when it is reflected off the boundaries. Thus, as opposed to periodic boundary conditions, only the total number of fermions is conserved
\begin{align} N=\sum_a\int_{0}^{L} dx \left(\psi^{\dagger}_{La}(x,t)\psi_{La}(x,t)+\psi^{\dagger}_{Ra}(x,t)\psi_{Ra}(x,t)\right).\label{number}\end{align}
We shall see in the next section that one can construct exact wavefunctions that are labelled by the above conserved quantities, that are the total number of particles $N$ and the total $z$ component of the spin $s^z$. 

\section{The generalized Bethe ansatz wavefunction}
In this section we follow the generalized Bethe ansatz framework \cite{PasnooriKondo} and construct exact solution to the time-dependent Schrodinger equation. We shall see that the consistency of the wavefunction imposes constraints on the bulk interaction strengths and as well as the time-dependent boundary conditions such that the system is integrable. Just as in the regular Bethe ansatz, we can construct wavefunctions that are labeled by the total number of fermions $\ket{N}$ and look for solutions to the time-dependent Schrodinger equation
\begin{align}i\partial_t\ket{N}=H\ket{N}. \label{SE}
\end{align}
Here $H$ stands for the Hamiltonian of both the $U(1)$ Thirring model and the $SU(2)$ Gross-Neveu model. As mentioned before, we shall demonstrate the method explicitly for the latter case and discuss the case of the $U(1)$ Thirring model in the end. 
\begin{center}
\begin{figure}
\includegraphics[width=0.75\columnwidth]{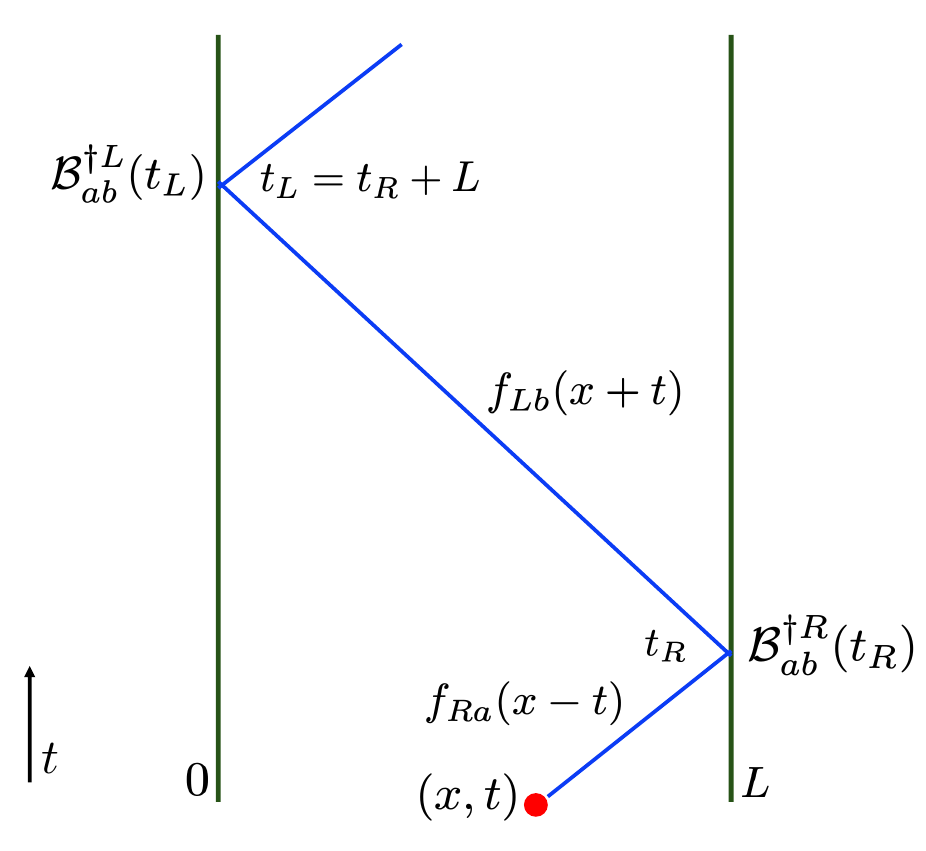}
\caption{The figure depicts the scattering of a right moving particle off the right boundary at time $t_R$ where it turns into a left mover. It then moves to the left boundary where it scatters off the left boundary and turns into a right mover at time $t_L=t_R+L$. The amplitudes of the particle just before and just after the scattering at the right boundary at time $t_R$ are related by the boundary S-matrix $-\mathcal{B}^{\dagger R}_{ab}(t_R)$.  Similarly, the amplitudes just before and after the scattering at the left boundary at time $t_L$ are related by the left boundary S-matrix by $-\mathcal{B}^{\dagger L}_{ab}(t_L)$. }
\label{fig:picture1}
\end{figure}
\end{center}

\subsection{One particle case}
Let us first consider the case of $N=1$, we have
\begin{align}\ket{1}\hspace{-0.5mm}=\hspace{-1mm}\sum_a\int_0^L \hspace{-1.5mm}dx \;\psi^{\dagger}_{Ra}(x,t)F_{Ra}(x,t)+\psi^{\dagger}_{La}(x,t)F_{L,a}(x,t)\ket{0}.\label{1pwf}
\end{align}
Using (\ref{1pwf}) in the Schrodinger equation (\ref{SE}), one obtains the following equation
\begin{align}\nonumber&-i\int_0^Ldx \;\psi^{\dagger}_{Ra}(x,t)(\partial_x+\partial_t)F_{Ra}(x,t)\\&+i\int_0^Ldx\; \psi^{\dagger}_{La}(x,t)(\partial_x-\partial_t)F_{La}(x,t)=0.\label{1pSE}
\end{align}
Since $F_{Ra}(x,t)$ and $F_{La}(x,t)$ correspond to the right and left moving particles, it is natural to take the following ansatz 
\begin{align} \nonumber
F_{Ra}(x,t)=f_{Ra}(z)\theta(x)\theta(L-x),\\
F_{La}(x,t)=f_{La}(\bar{z})\theta(x)\theta(L-x).\label{1pansatz}
\end{align}
Here $z=x-t,\bar{z}=x+t$ and $\theta(x)$ is the Heaviside step function with the convention $\theta(x)=1$ for $x>0$, $\theta(x)=0$ for $x<0$  and $\theta(0)=1/2$.
Using (\ref{1pansatz}) in (\ref{1pSE}), one can see that the derivatives acting on $f_{Ra}(z)$ and $f_{La}(\bar{z})$ cancel trivially. The derivatives acting on the Heaviside functions gives rise to boundary terms, which upon using the boundary conditions (\ref{OBC}) yield
\begin{align}f_{La}(L+t_R)&=-\mathcal{B}^{\dagger R}_{ab}(t_R)f_{Rb}(L-t_R),\\f_{Ra}(-t_L)&=-\mathcal{B}^{\dagger L}(t_L)_{ab} f_{Lb}(t_L).
\end{align}
The first equation describes the scattering of the right moving particle against the right boundary where it turns into a left mover at time $t_R$ and relates these two amplitudes in its vicinity $x=L$. See Fig (\ref{fig:picture1}). Similarly, the second equation describes the scattering of the left moving particle against the left boundary where it turns into a right mover at time $t_L=t_R+L$ and relates the two amplitudes near its vicinity $x=0$. Using the above two equations, one obtains the following matrix difference equations which constrain the amplitudes $f_{Ra}(z)$ and $f_{La}(\bar{z})$ for a given set of time-dependent boundary conditions
\begin{align} &f^{R}(z-2L)=\mathcal{B}^{\dagger L}(2L-z)\mathcal{B}^{\dagger R}(L-z)f^R(z),\\
& f^{L}(\bar{z}+2L)=\mathcal{B}^{\dagger L}(\bar{z}+L)\mathcal{B}^{\dagger R}(\bar{z})f^L(\bar{z}).\end{align}

The above matrix difference equations do not impose any constraints on the boundary conditions. This is expected since the one particle sector is trivially `integrable' in any system. For a given set of boundary conditions, one can solve the above matrix difference equations to obtain the explicit form of the amplitudes. We shall not elaborate on this further and move on to the two particle sector which is non trivial, and as we shall see, integrability imposes constraints on the time-dependent boundary conditions.

\subsection{Two particle case}
Now let us consider the case of two particles, we have
\begin{align} \ket{2}=\sum_{\chi_i,a_i}\prod_{i=1}^2\int_0^Ldx_i \psi^{\dagger}_{\chi_ia_i}(x_i,t)\mathcal{A}F^{\chi_1\chi_2}_{a_1a_2}(x_1,x_2,t)\ket{0}.\label{2pwf}
\end{align}
Here $\mathcal{A}$ is the anti-symmetrization symbol and $\chi_i\in L,R$, $i=1,2$ represents the chirality. $a_i=\uparrow,\downarrow$ where $i=1,2$ represent the spin. Since the particles interact with each other, one needs to distinguish between different ordering of the particles in the configuration space.  Just as in the periodic boundary conditions case, even though the particles with same chiralities do not interact, integrability requires one to distinguish between different ordering of the particles with the same chirality as well. We take the following ansatz
\begin{align}\nonumber F^{\chi_1\chi_2}_{a_1a_2}(x_1,x_2,t)=\big(f^{\chi_1\chi_2,12}_{a_1a_2}(z^{\chi_1}_1,z^{\chi_2}_2)\theta(x_2-x_1)\\\nonumber +f^{\chi_1\chi_2,21}_{a_1a_2}(z^{\chi_1}_1,z^{\chi_2}_2)\theta(x_1-x_2)\big)\\\times \theta(x_1)\theta(L-x_1)\theta(x_2)\theta(L-x_2).\label{2pansatz}
\end{align}
Here with a slight abuse of notation we have used $z^{R}\equiv z$ and $z^{L}\equiv \bar{z}$. The first superscript denotes the chiralities whereas the second superscript denotes the position ordering of the particles.
Using (\ref{2pwf}) and (\ref{2pansatz}) in (\ref{SE}), one obtains the following set of equations. From here on we will suppress the spin indices for the ease of notation, unless otherwise required.

\begin{align} \label{2peq1}f^{RL,21}(z,\bar{z})=(2i-g(t)\vec{A})^{-1}(2i+g(t)\vec{A})f^{RL,12}(z,\bar{z})\\f^{LR,12}(\bar{z},z)=(2i-g(t)\vec{A})^{-1}(2i+g(t)\vec{A})f^{LR,21}(\bar{z},z).\label{2peq2}
\end{align}
 In addition one obtains boundary terms which we shall discuss shortly. As we shall see below, the above equations (\ref{2peq1}) and (\ref{2peq2}) which relate the amplitudes with different ordering of the particles with opposite chiralities give rise to the bulk S-matrix which exchanges these particles. In the above equations, $(X)^{-1}$ denotes the inverse of the operator $X$. We make the series of transformations
 \begin{align} t= t^*+\Delta x,\;\; x=(x_1+x_2)/2, \;\;\Delta x=(x_2-x_1)/2\end{align} and use $z=x_1-t^*, \bar{z}_2=x_2+t^*$ in the equation (\ref{2peq1}). We then take the inverse of $(2i-g(t)\vec{A})$ and simplify the resulting expression, which yields

\begin{align} \nonumber f^{RL,21}(z_1,\bar{z}_2)=S_{12}(\bar{z}_2-z_1)f^{RL,12}(z_1,\bar{z}_2)\\S_{12}(\bar{z}_2-z_1)=R_{12}(f(\bar{z}_2-z_1))e^{i\phi(\bar{z}_2-z_1)},\label{2psmat}
\end{align}
where $R_{ij}(\lambda)$ is just the XXX R-matrix:
\begin{align}R_{ij}(\lambda)=\frac{i \lambda I_{12}+P_{12}}{i\lambda+1}.\label{XXXRmat}\end{align}
Here $I_{12}$ is the identity and $P_{12}$ is the permutation operator \cite{PasnooriKondo} which is acting in the spin spaces of the particles 1 and 2. $e^{i\phi(x)}$ is a factor which is not relevant for our current discussion and we shall omit this in the following.  As mentioned above, the function $f(t)$ is related to the bulk interaction strength through a non universal relation $f(t)=\frac{1}{2g(t/2)}\left(1-\frac{3g(t/2)^2}{4}\right)$. Different regularization schemes involving the definition of the Heaviside function $\theta(x)$ give rise to different relations (\ref{gnrelr}), but they all coincide in the universal regime which corresponds to small values of the coupling strengths \cite{Pasnoorigrossneveu,PasnooriRG}. Integrability requires that the function $f(t)$ is linear in time, which corresponds to the following form of the bulk interaction strength $g(t)$ in the universal regime
\begin{align}g(t)=\frac{1}{4(\alpha t+\beta/2)}, \;\; \alpha,\beta\;\;  \text{are constants}.\label{intstrength}
\end{align}
Here we would like to note that one does not have to a priori work with an integrable bulk interaction strength. One can instead work with the most general time dependent bulk interaction strengths and boundary conditions. In the case of periodic boundary conditions, integrability imposes constraints on the bulk interaction strength in the three particle sector (with at least one particle having a different chirality), which is imposed by the Yang-Baxter equation. In the case of open boundary conditions, the integrable bulk and boundary conditions are fixed in the two particle sector when one requires that the resulting bulk and boundary S-matrices satisfy the reflection equation, as we shall see in the following. For the simplicity of the arguments, we shall proceed forward with the integrable bulk interaction strength (\ref{intstrength}) and derive the boundary conditions compatible with integrability.

\begin{center}
\begin{figure}
\includegraphics[width=1.0\columnwidth]{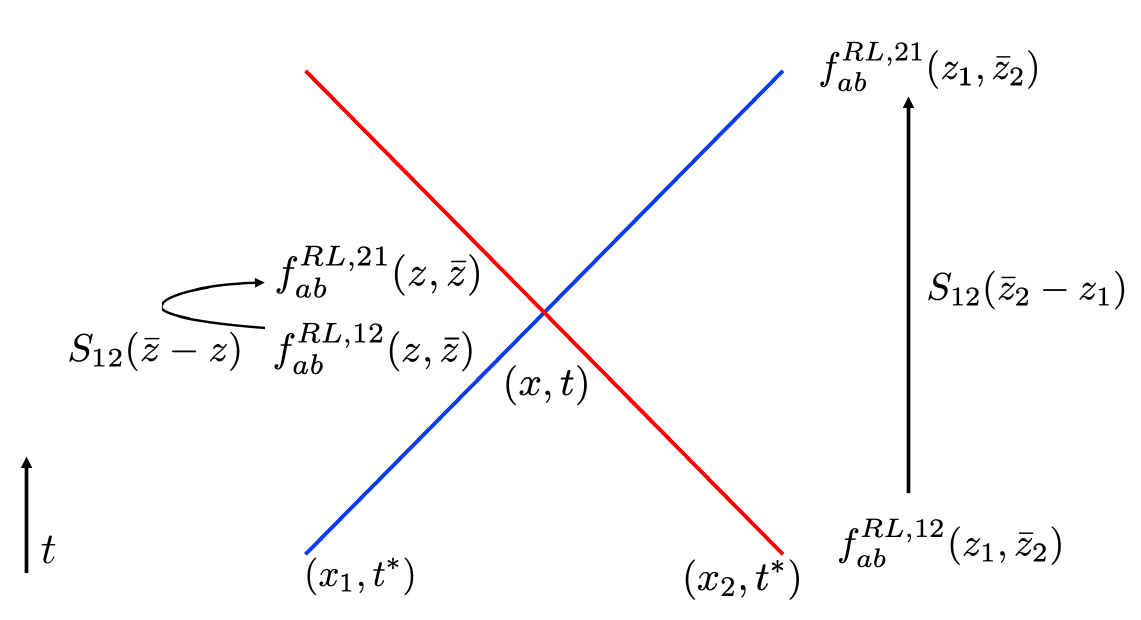}
\caption{The figure depicts the scattering of a right and a left moving particle. At time $t^*$ they are located at spatial coordinates $x_1$ and $x_2$. They move toward each other and meet at the point $x=(x_1+x_2)/2$ at time $t=t^*+\Delta x$, where $\Delta x=(x_1-x_2)/2$. The amplitudes right before and right after the scattering $f^{RL,12}(z,\bar{z})$, $f^{RL,21}(z,\bar{z})$ are related to each other through the relation (\ref{2peq1}). The relation between the amplitude before the collision $f^{RL,12}(z_1,\bar{z}_2)$ and the amplitude after the collision $f^{RL,21}(z_1,\bar{z}_2)$ are related through the S-matrix $S_{12}(\bar{z}_2-z_1)$ (\ref{2psmat}). }
\label{fig:picture2}
\end{figure}
\end{center}

Before proceeding further let us gain some physical intuition behind the transformations and the resulting relations (\ref{2psmat}) obtained above. Consider the amplitude $f^{RL,12}(z_1,\bar{z}_2)$ where at time $t^*$, the right moving particle 1 is on the left side of the left moving particle 2 as shown in Fig (\ref{fig:picture2}). The two particles approach each other and interact at the spatial coordinate $x=(x_1+x_2)/2$ at time $t=t^*+\Delta x$. Thus the equation (\ref{2peq1}) is relating the two amplitudes at the point in spacetime where the interaction has occurred. To relate the amplitudes at time $t^*$, we make the transformations discussed above which simply transport the particles backwards in time. Note that in the resulting equation (\ref{2psmat}), the bulk S-matrix corresponds to the interaction between the particles at time $t=t^*+\Delta x= (\bar{z}_2-z_1)/2$. The above analysis provides the rule of thumb where we choose the relation between two amplitudes such that it is fixed by the S-matrix with the interaction strength corresponding to the actual time where they interact physically.  Note that the amplitude $f^{RL,21}(z_1,\bar{z}_2)$ is unphysical for $t^*<t$, but nevertheless, its functional form is fixed by the relation (\ref{2psmat}), where it is related to the physical amplitude $f^{RL,12}(z_1,\bar{z}_2)$. In contrary, for $t^*>t$, the amplitude $f^{RL,21}(z_1,\bar{z}_2)$ is physical whereas the amplitude $f^{RL,12}(z_1,\bar{z}_2)$ is unphysical and they both are again related through the relation (\ref{2psmat}). For additional details, we refer the reader to the supplementary metarial of \cite{PasnooriKondo}. To summarize, the two particle S-matrix (\ref{2psmat}) relates amplitudes before and after collision, where one amplitude is physical whereas the other is unphysical. The physical or unphysical nature of the amplitude is dictated based on whether one chooses the time in the S-matrix to be greater than or smaller than the actual time of the scattering. When the time in the S-matrix is chosen to be exactly equal to the time of the scattering event, then both the amplitudes related by the S-matrix are physical. This structure is common to all the bulk and boundary S-matrices, and we shall not elaborate on this further.

Similarly, by using the same set of transformations but with $x_1\leftrightarrow x_2$, one obtains the following relation
\begin{align}  f^{LR,12}(\bar{z}_1,z_2)=S_{12}(\bar{z}_1-z_2)f^{LR,21}(\bar{z}_1,z_2),\label{2psmat2}
\end{align}
where the S-matrix takes the same form as in (\ref{2psmat}). In addition to the relations (\ref{2peq1}) and(\ref{2peq2}) as mentioned above, one obtains the following boundary terms corresponding to the left boundary

\begin{align} \label{2pb1}f^{RR,12}(-t,x_2-t)&=-\mathcal{B}_1^{\dagger L}(t)f^{LR,12}(t,x_2-t),\\\label{2pb3}f^{RR,21}(x_1-t,-t)&=-\mathcal{B}_2^{\dagger L}(t)f^{RL,21}(x_1-t,t),\\\label{2pb5}f^{RL,12}(-t,x_2+t)&=-\mathcal{B}^{\dagger L}_1(t)f^{LL,12}(t,x_2+t),\\\label{2pb6}f^{LR,21}(x_1+t,-t)&=-\mathcal{B}^{\dagger L}_2(t) f^{LL,21}(x_1+t,t).\end{align}

Similarly, one obtains the following boundary terms corresponding to the right boundary
\begin{align}&\label{2pb2}f^{LR,21}(L+t,x_2-t)=-\mathcal{B}_1^{\dagger R}(t)f^{RR,21}(L-t,x_2-t),\\&\label{2pb4}f^{RL,12}(x_1-t,L+t)=-\mathcal{B}_2^{\dagger R}(t)f^{RR,12}(x_1-t,L-t),\\&\label{2pb7}f^{LL,21}(L+t,x_2+t)=-\mathcal{B}^{\dagger R}(t)f^{RL,21}(L-t,x_2+t),\\&\label{2pb8}f^{LL,12}(x_1+t,L+t)=-\mathcal{B}^{\dagger R}_2(t) f^{LR,12}(x_1+t,L-t).\end{align}

\begin{center}
\begin{figure}
\includegraphics[width=0.61\columnwidth]{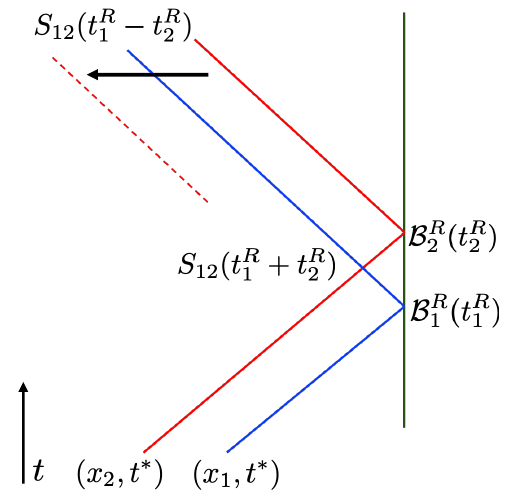}
\caption{Two right moving particles at spacetime coordinates $(x_1,t^*)$ and $(x_2,t^*)$ scatter off the right boundary at times $t_1^R$ and $t_2^R$ respectively, and the turn into left movers. The S-matrices associated with their reflection at the right boundary are $\mathcal{B}^{\dagger R}_1(t_1^R)$ and $\mathcal{B}^{\dagger R}_2(t_2^R)$ respectively. After the particle 1 has scattered off the right boundary, it interacts with the particle 2 with the S-matrix $S_{12}(t_1^R+t_2^R)$. After the second particle has scattered with the right boundary, the two particles can be exchanged with the corresponding S-matrix $S_{12}(t_1^R-t_2^R)$. }
\label{fig:picture3}
\end{figure}
\end{center}

Consider the relation (\ref{2pb2}). This relation describes the scattering process at time $t$ where the particle 1 moving to the right interacts with the right boundary and turns into a left mover, while the particle 2 is a right mover which is located on the left side of particle 1. The equation relates these amplitudes in the vicinity of the right boundary. Consider the amplitude before the collision at time $t^*$ where the right moving particle is located at position $x_1$ at shown in Fig (\ref{fig:picture3}). This amplitude can be related to the amplitude where the particle 1 has scattered off the right boundary and turned into a left mover by following the rule of thumb used in the case of periodic boundary conditions discussed above, where the two amplitudes should be related by an S-matrix with interaction strength corresponding to the time where the scattering occurs physically. In this case, this corresponds to time $t=t^*+L-x_1$. Thus using this transformation in (\ref{2pb2}), we obtain 
\begin{align}\nonumber&f^{LR,21}(t+2L-x_1,x_2-t)=\\&-\mathcal{B}^{\dagger R}_1(t+L-x_1)f^{RR,21}(x_1-t,x_2-t).\label{rel1}
\end{align}
Note that in addition to the shift in time, we have to shift the spatial coordinate of the particle 2: $x_2\rightarrow x_2+L-x_1$ to obtain the above relation. This results in the amplitude on the left side in the above equation being unphysical whereas the amplitude on the right side is physical. Note that we have chosen the time in the boundary S-matrix to be equal to the time $t^*<t_1^R$ which corresponds to the time before the scattering of the particle 1 with the right boundary. If we instead choose the time in the boundary S-matrix too be greater than $t_1^R$, then the amplitude on the left side of the above equation would be physical, whereas, the amplitude on the right side is unphysical. Note that this situation is similar to the case of the scattering between particles 1 and 2 (\ref{2psmat}). Similarly, if we choose the time in the boundary S-matrix to be the exact time of the scattering event, then both the amplitudes related by the boundary S-matrix are physical.

Now let us continue the discussion of the scattering events near the right boundary. After the particle 1 scatters off the right boundary and turns into a left mover, we obtain the amplitude $f^{LR,21}(t+2L-x_1,x_2-t)$ (\ref{rel1}). The particle 1 then interacts with the particle 2 which is a right mover and they exchange their positions which results in the amplitude $f^{LR,12}(t+2L-x_1,x_2-t)$. These two amplitudes can be related to each other by using the transformation $x_1\rightarrow2L-x_1$ in (\ref{2psmat2}). The particle 2 then moves to the right and interacts with the right boundary 
which is described by the equation (\ref{2pb8}). From now on we use the rule of thumb mentioned above without explicitly stating it. We make the transformation $t\rightarrow t+L-x_2, x_1\rightarrow L+x_2-x_1$ in (\ref{2pb8}), and obtain
\begin{align}\nonumber&f^{LL,12}(t+2L-x_1,t+2L-x_2)\\&=-\mathcal{B}^{\dagger R}_2(t+L-x_2)f^{LR,12}(t+2L-x_1,x_2-t).
\end{align}

Now consider the amplitude $f^{LL,12}(t+2L-x_1,t+2L-x_2)$. The particle 2 which is a left mover can be moved past the particle 1 which is also a left mover which results in the amplitude $f^{LL,21}(t+2L-x_1,t+2L-x_2)$. Note that just as in the case of periodic boundary conditions, these amplitudes are not related to each other by the Hamiltonian. One should choose the S-matrix relating these amplitudes such that the S-matrices satisfy Yang-Baxter equation. In the current case, they also have to satisfy the reflection equation which shall be discussed shortly. The S-matrix which is consistent with these equations is same as that in the case of periodic boundary conditions, and it is given by 
\begin{align}\nonumber& f^{LL,21}(t+2L-x_1,t+2L-x_2)\\&=S_{12}(\bar{z}_1-\bar{z}_2)f^{LL,12}(t+2L-x_1,t+2L-x_2),\label{rel2}\end{align}
Here we note that the rule of thumb used for the S-matrix corresponding to the two particles with opposite chiralities does not work for the case of two particle with the same chiralities, since they never physically meet. The rule of thumb for such an S-matrix (\ref{rel2}) between the particles with the same chiralities is to choose its `argument' such that is equal to the distance between the particles.

\begin{center}
\begin{figure}
\includegraphics[width=0.65\columnwidth]{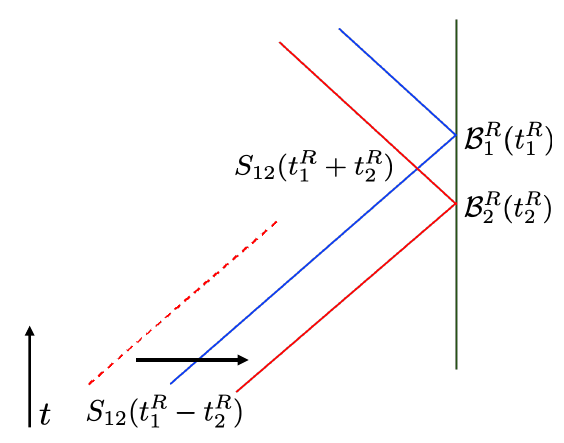}
\caption{The positions of the two right moving particles can be exchanged whose corresponding S-matrix is $S_{12}(t_1^R-t_2^R)$. The particles then scatter off the right boundary at times $t_2^R$ and $t_1^R$ respectively, and the turn into left movers. The S-matrices associated with their reflection at the right boundary are $\mathcal{B}^{\dagger R}_2(t_2^R)$ and $\mathcal{B}^{\dagger R}_1(t_1^R)$ respectively. After the particle 2 has interacted with the right boundary, it then interacts with the particle 1 with the S-matrix $S_{12}(t_1^R+t_2^R)$. }
\label{fig:picture4}
\end{figure}
\end{center}

Now starting with the amplitude $f^{RR,21}(x_1-t,x_2-t)$, consider the above processes happening in the reverse order as depicted in Fig (\ref{fig:picture4}), that is, the particle 2 which is a right mover moves past particle 1. Just as in the above case, these amplitudes are not related to each other by the Hamiltonian. One should choose the S-matrix relating these amplitudes such that the S-matrices satisfy Yang-Baxter equation. We have
\begin{align} f^{RR,12}(x_1-t,x_2-t)=S_{12}(z_1-z_2)f^{RR,21}(x_1-t,x_2-t),\label{rel3}\end{align}
Note that this S-matrix is exactly same as that corresponding to the exchange of the two right moving particles (\ref{rel2}), since $z_1-z_2=\bar{z}_1-\bar{z}_2$. The particle 2 then interacts with the right boundary and turns into a left mover. The particles 1 and 2 then interact and exchange positions. Then particle 1 interacts with the right boundary and turns into a left mover. We end up with the same amplitude as above which is  $f^{LL,21}(t+2L-x_1,t+2L-x_2)$. For the wavefunction to be consistent, both these processes should be equivalent, which gives rise to the reflection equation corresponding to the right boundary
\begin{align}\nonumber &
S_{12}(t_1^R-t_2^R)\mathcal{B}_1^R(t_1^R)S_{12}(t_1^R+t_2^R)\mathcal{B}^R_2(t_2^R)\\&=\mathcal{B}^R_2(t_2^R)S_{12}(t_1^R+t_2^R)\mathcal{B}_1^R(t_1^R)S_{12}(t_1^R-t_2^R),\label{reflectionR}
\end{align}
where we have used the notation $t_i^R=t+L-x_i$, $i=1,2$.

Now one can proceed forward and consider the processes where the two particles scatter off the left boundary. Following the same procedure as above, one obtains the reflection equation corresponding to the left boundary
\begin{align}\nonumber &
S_{12}(t_1^L-t_2^L)\mathcal{B}_1^L(t_1^L)S_{12}(t_1^L+t_2^L)\mathcal{B}^L_2(t_2^L)\\&=\mathcal{B}^L_2(t_2^L)S_{12}(t_1^L+t_2^L)\mathcal{B}_1^L(t_1^L)S_{12}(t_1^L-t_2^L),\label{reflectionL}
\end{align}
where $t_i^L=t_i^R+L$. The occurrence of the shift in the time is due to the fact that it takes time $t_i^L-t_i^R=L$ for the particle $i$ to move to the left boundary after it has scattered off the right boundary. The reflection equations have been solved by Cherednik \cite{Cherednik1} whose solutions are given by the $K$ matrices \footnote{Note that the two reflections equations are the same unlike in the eigenvalue problem involving the transfer matrix, where one obtains dual reflection equation for the left boundary (or vice versa). We shall comment on this elsewhere.}
\begin{align}\nonumber& \mathcal{B}^{\dagger R}(\lambda)=K^-(\lambda)=\xi^R+i\lambda \sigma^z,\\ \;&\mathcal{B}^{\dagger L}(\lambda)=K^+(\lambda)=\xi^L+i\lambda\sigma^z.\label{XXXkmat}
\end{align}
Thus we have derived the most general time dependent boundary conditions preserving the $U(1)$ symmetry (which corresponds to the conservation of total $z$ component of the spin) for the system to be integrable. Below we provide their form which shows their explicit dependence on time 
\begin{align}&\label{XXXbcl}\psi_{La}(0,t)=-\mathcal{B}^{L}_{ab}(t)\psi_{Rb}(0,t), \\ \label{XXXbcr} &\psi_{Ra}(L,t)=-\mathcal{B}^{R}_{ab}(t)\psi_{Lb}(L,t). \\
 &\mathcal{B}^{\dagger \chi}(t)=\xi^{\chi}+if(t) \sigma^z, \;\; \chi=L,R. \label{XXXbc}\end{align}

Here we note that the parameters $\xi^L$ and $\xi^R$ appearing in the above $K$ matrices correspond to RG-invariants in the static model \cite{PAA2,PAA1}, where their values determine the boundary phases exhibited by the system, and in addition, they characterize the nature of the boundary bound states and their energies. In the current case where the Hamiltonian is time-dependent, they are dynamical invariants. One naturally expects this correspondence between RG invariants and dynamical invariants \cite{pasnooriNHKtime} from the time-dependent integrability-RG flow connection \cite{Benhoare,PasnooriRG,pasnoorigrossneveu2}.

 We now have all the necessary relations to construct the transport operator which takes the particles around the system once. Starting from the amplitude $f^{RR,21}(x_1-t,x_2-t)$ Using (\ref{rel1}), (\ref{rel2}), (\ref{2psmat2}), one obtains
\begin{align}\nonumber&
f^{RR,12}(t-x_1-2L,x_2-t)=S_{12}(t_1^R-t_2^R+2L)\\&\mathcal{B}^{\dagger L}_1(t_1^R+L) S_{12}(t_1^R+t_2^R)\mathcal{B}^{\dagger R}_1(t_1^R)f^{RR,21}(x_1-t,x_2-t)\label{2pBqKZ1}.\end{align} Similarly transporting particle 2, we have
\begin{align}&\nonumber
 f^{RR}(x_1-t,x_2-t-2L)=\mathcal{B}^L_2(t_2^R+L)\\&S_{12}(t_1^R+t_2^R)\mathcal{B}_2^R(t_2^R)S_{12}(t_2^R-t_1^R) f^{RR,21}(x_1-t,x_2-t).  \label{2pBqKZ2}
\end{align}
Thus we obtain the two matrix difference equations (\ref{2pBqKZ1}) and (\ref{2pBqKZ2}) which are the constraints imposed by integrability on the amplitude $f^{RR,21}(x_1-t,x_2-t)$. One can interpret the transport operators as relating the amplitudes at times $t$ and $t+2L$, which is the time it takes for a particle to move around the system once. 

\subsection{$N$ particle case}
In the case of two particles $N=2$, we have seen that there exist two bulk S-matrices, where one corresponds to the scattering of two particles with same chiralities and the other one corresponds to the scattering of two particles with opposite chiralities. In addition, there exists two boundary S-matrices, each corresponding to the scattering of one particle with the respective boundary. These bulk and boundary S-matrices satisfy the reflection equations. In the case of $N>2$, there exist multiple scattering events between particles with the same and different chiralities in the bulk. These S-matrices satisfy the Yang-Baxter equation similar to the case of periodic boundary conditions. Below we shall briefly discuss the scattering processes in the case of three particles $N=3$ and refer the reader to \cite{Pasnoorigrossneveu} for detailed exposition. 

Consider the case of three particles $N=3$. There exist in total of 48 amplitudes: These can be grouped into eight sets of amplitudes, where each set corresponding to a particular combination of the chiralities of the particles. Within each set, there exist 6 different amplitudes each corresponding to a specific ordering of the particles with respect to each other. For a particular chirality of a specific particle in the bulk, there exist sixteen amplitudes, eight at each boundary, corresponding to the scattering of the two other particles with the respective boundary and among themselves. These are described by the reflection equations (\ref{reflectionL}) and (\ref{reflectionR}) discussed in the previous subsection. In addition to these scattering events associated with the boundaries, there exists scattering between the three particles. These events can be categorized into four groups depending on the chiralities of the particles: \textit{a)Three right movers b)Three left movers c)Two right movers and one left mover d)One right over and two left movers}.

\begin{center}
\begin{figure}[h]
\includegraphics[width=0.75\columnwidth]{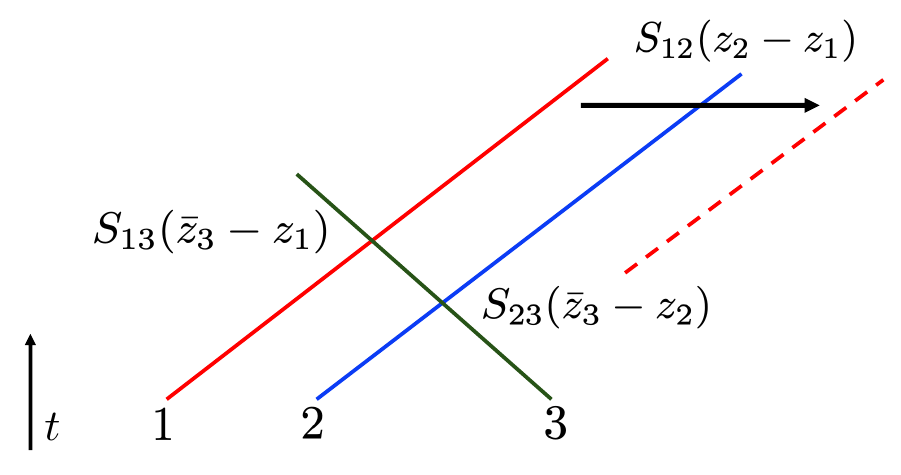}
\caption{The figure depicts two right moving particles 1 and 2 scattering with a left moving particle 3. The particle 2 is on the right side of particle 1, and hence interacts with particle 3 first with the S-matrix $S_{23}(\bar{z}_3-z_2)$. The particle 1 then interacts with particle 3 with an S-matrix $S_{13}(\bar{z}_3-z_1)$. The particle 1 is then moved past particle 2 where the corresponding S-matrix is $S_{12}(z_2-z_1)$. }
\label{fig:picture5}
\end{figure}
\end{center}

Let us consider the case of two right moving particles and one left moving particle as shown in Fig (\ref{fig:picture5}). The right moving particle 2 is to the right side of particle 1 and hence it interacts with the left moving particle 3 first. The S-matrix of this scattering is $S_{23}(\bar{z}_3-z_2)$, which takes the same form as in the two particle case discussed in the previous subsection (\ref{2psmat}). Next, the particle 1 and particle 3 approach each other and interact with an S-matrix $S_{13}(\bar{z}_3-z_1)$. After these two scattering events, the order of the particles 1 and 2 remains the same where particle 2 is on the right side of particle 1. One can exchange their positions such that particle 1 is on the right side of particle 2. The S-matrix associated with this scattering of the two particles with same chiralities takes the same form as that in the two particle case: $S_{12}(z_2-z_1)$. 

\begin{center}
\begin{figure}[h]
\includegraphics[width=0.75\columnwidth]{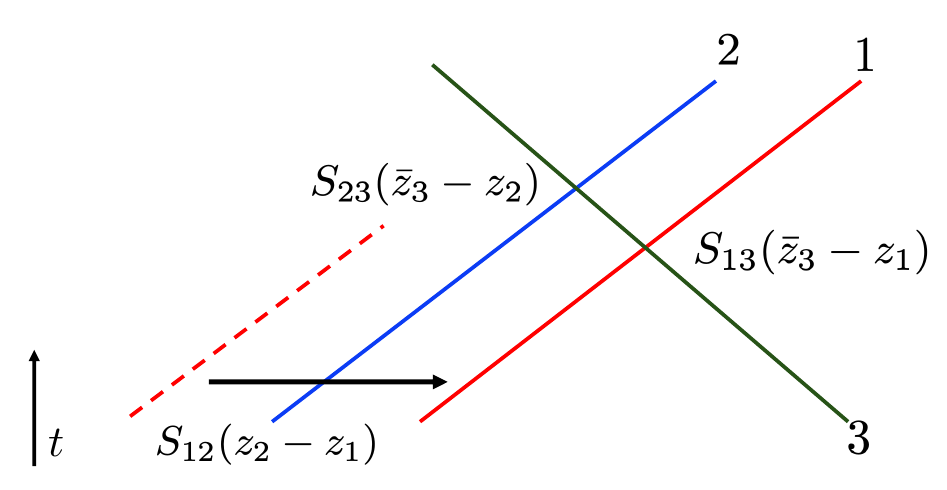}
\label{fig:picture6}
\caption{The figure depicts two right moving particles 1 and 2 scattering with a left moving particle 3. The particle 1 is on the left side of particle 2. It is moved to the right side of particle 2 where the corresponding S-matrix is $S_{12}(z_2-z_1)$. The particle 1 then interacts with particle 3 first with the S-matrix $S_{13}(\bar{z}_3-z_1)$. The particle 2 then interacts with particle 3 with an S-matrix $S_{23}(\bar{z}_3-z_2)$. The events in this figure occur exactly in the opposite order compared to fig (\ref{fig:picture5}).  }
\end{figure}
\end{center}

One can arrive at the same final amplitude by considering the above scattering events occurring in the reverse order as shown in Fig (\ref{fig:picture6}): The particle 1 and 2 are exchanged with the S-matrix $S_{12}(z_2-z_1)$. The particle 1 then interacts with the particle 3 with an S-matrix $S_{13}(\bar{z}_3-z_1)$ followed by particle 2 interacting with particle 3 with an S-matrix  $S_{23}(\bar{z}_3-z_2)$. Since the initial and final amplitudes are the same, the two different orders of the scattering events described above should be equivalent. This simply corresponds to the Yang-Baxter equation:
\begin{align}\nonumber &S_{12}(z_2-z_1)S_{13}(\bar{z}_3-z_1)S_{23}(\bar{z}_3-z_2)\\&=S_{23}(\bar{z}_3-z_2)S_{13}(\bar{z}_3-z_1)S_{12}(z_2-z_1). \label {YB1}\end{align}
The case of two left moving particles and one right moving particle is similar to the case discussed above. If we consider particle 1 and 2 to be left movers and 3 to a right mover, then the resulting Yang-Baxter equations can be obtained from the above (\ref{YB1}) with the transformations: $z_1 \rightarrow \bar{z}_1$, $z_2\rightarrow \bar{z}_2$ and $\bar{z}_3\rightarrow z_3$. 

Now consider the case where all three particles have the same chiralities, say right movers. Starting from an amplitude with some specific ordering of the particles where particle 1 is to left side of particle 2, which is on the left side of particle 3, one can exchange the particles with each other in two different ways and arrive at the same final amplitude. This gives rise to the following Yang-Baxter equation:
\begin{align}\nonumber &S_{12}(z_2-z_1)S_{13}(z_3-z_1)S_{23}(z_3-z_2)\\&=S_{23}(z_3-z_2)S_{13}(z_3-z_1)S_{12}(z_2-z_1) \label {YB2}.\end{align}
The case where all the three particles are left movers is similar and the resulting Yang-Baxter equation can be obtained from the above using the transformation: $z_i\rightarrow \bar{z}_i$, $i=1,2,3$. 
The case of three particles exhausts all the possible bulk and boundary S-matrices and the resulting Yang-Baxter and reflection equations associated with the model. Similar to the case of two particles, starting from one particular amplitude, one can construct transport operators which transport a particle around the system once. One obtains a matrix difference equation for the transportation of each particle similar to the case of two particles. We shall not elaborate on this further and consider the case of general $N$.

For the case of general $N$, we have 

\begin{align}
\ket{N}=\sum_{\chi_j,\sigma_j}\prod_{j=1}^{N}\int_{0}^L\hspace{-2.5mm}dx_j \psi^{\dagger}_{R\sigma_j}(x_j)\mathcal{A}F^{\{\chi_j\}}_{\{\sigma_i\}}(x_1,...,x_N,t)\ket{0}.\label{npwf1}
\end{align}
Here, $\sigma_1...\sigma_N\equiv\{\sigma_i\}$ denote the spin and $\chi_1...\chi_N\equiv\{\chi_i\}$ denote the chiralities of the particles. The ansatz wavefunction $F^{\{\chi_j\}}_{\{\sigma_i\}}(x_1,...,x_N,t)$ takes the following form
\begin{align} 
F^{\{\chi_i\}}_{\{\sigma_i\}}\hspace{-0.3mm}(x_1,..,x_N,t)
\hspace{-1mm}= \hspace{-1.3mm}\sum_Q \theta(\{x_{Q(j)}\})  f^{\{\chi_i\},Q}_{\{\sigma_i\}}(x_1,...,x_N,t).\label{npwf2}
\end{align}

Here $Q$ denotes a permutation of the position orderings of particles and  $\theta(\{x_{Q(j)}\})$ is the Heaviside function that vanishes unless $x_{Q(1)} \le \dots \le x_{Q(N)}$. Here $f^{\{\chi_i\},Q}_{\{\sigma_i\}} (x_1,...,x_N,t)$ is the amplitude corresponding to the ordering of the particles denoted by $Q$. Just as in the case of two and three particles, the scattering of two particles is associated with a bulk S-matrix. The amplitudes corresponding to different ordering of the particles are related to each other through the various bulk S-matrices and they satisfy the Yang-Baxter equations (\ref{YB1}) and (\ref{YB2}). The amplitudes which correspond to different chirality of the particle near the boundary are related to each other through the boundary S-matrices associated with the respective boundary. The bulk S-matrices corresponding to the scattering of two particles with each other and the boundary S-matrices corresponding to the scattering of the particles with the respective boundary satisfy the reflection equations (\ref{reflectionL}) and (\ref{reflectionR}). Similar to the case of two particles, starting from a reference amplitude, one can construct a transport operator which transports the particle around the system once. This results in a matrix difference equation, which we shall discuss below.

Consider the amplitude $f^{\{R\},N...1}_{\{\sigma_j\}}(x_1,...,x_N,t)$ where all the particles are right movers and ordered such that $x_N<x_{N-1}<...<x_1$. Let us move the particle $j$ to the right. It first interacts with the particle $j-1$. Using the rule of thumb for the S-matrix between two particles with the same chirality, one obtains the S-matrix for this scattering: $S_{ij}(z_{j-1}-z_j)$. Similarly, the particle $j$ can be moved past all the particle $j-2,...1$ with the corresponding S-matrices. The particle $j$ then reflects off the right boundary and turns into a left mover. To find the boundary S-matrix we use the rule of thumb, where on chooses the S-matrix such that it corresponds to the actual time of the scattering event, which corresponds to $t_j^R=t+L-x_j$, and thus we obtain $K^{-}_j(t_j^R)$. The particle $j$ moves to the left and scatters with the particle 1. Again using the rule of thumb, where we choose the S-matrix between particles with opposite chiralities such that it corresponds to the actual time of the scattering event, we thus have $S_{j1}(t_j^R+t_1^R)$. The particle $j$ then scatters off the particles $2,...,N$ with the corresponding S-matrices. The particle $j$ then scatters off the left boundary and turns into a right mover. This physically occurs at time $t_j^R+L$, thus using the rule of thumb, we have the S-matrix: $K^{+}_j(t_j^R+L)$. The particle $j$ then can be moved past particle $N$. Note that the distance the particle $j$ has travelled is $2L-x_j+x_N$, thus using the rule of thumb for the S-matrix between two particles with the same chirality we have $S_{jN}(t_j^R-t_N^R+2L)$. Similarly it can be moved past the particles $N-1,...,j+1$ with the corresponding S-matrices. We thus obtain the following matrix difference equations 
\begin{align}\nonumber &f^{\{R\},N...1}(x_1-t,...,x_j-t-2L,...x_N-t)\\&= Z_j(t_1^R,...,t_N^R)\;  f^{\{R\},N...1}(x_1-t,...,x_j-t,...x_N-t),\label{BqKZ1}
\end{align}
where the transport operator $Z_j(t_1^R,...,t_N^R)$ which transports particle $j$ around the system once is given by
\begin{widetext}
\begin{align}\nonumber
 Z_j(t_i^R,..,t_N^R)=S_{jj+1}(t_j^R-t_{j+1}^R+2L)...S_{jN}(t_j^R-t_N^R+2L)K^{+}_j(t_j^R+L)S_{jN}(t_j^R+t_N^R)...S_{jj+1}(t_{j+1}^R+t_j^R)\\\times S_{j-1j}(t_{j-1}^R+t_j^R)...S_{j1}(t_j^R+t_1^R)K^{-}_j(t_j^R)S_{j1}(t_j^R-t_1^R)...S_{jj-1}(t_j^R-t_{j-1}^R).  \label{BqKZ2}
\end{align}
\end{widetext}
The equations (\ref{BqKZ1}), (\ref{BqKZ2}) are nothing but the boundary quantum Knizhnik-Zamolodchikov equations \cite{RishetikhinbqKZ}, which was already alluded in \cite{Pasnoorigrossneveu}. For the consistency of the wavefunction, transporting  particle `$i$' around the system once and then particle `$j$' around the system later should be equivalent to transporting them in the reverse order. This results in the following relation
\begin{align}\nonumber Z_j(t_i^R,...,t_i^R+2L,...,t_N^R)Z_i(t_1^R,...,t_N^R)=\\Z_i(t_i^R,...,t_j^R+2L,...,t_N^R)Z_j(t_1^R,...,t_N^R).\label{zerocurvature}
\end{align}
Note that in the above equations (\ref{BqKZ2}), the S-matrices are related to the rational XXX R-matrix, and thus (\ref{BqKZ1}) are the bqKZ equations corresponding to the rational XXX R-matrix. As mentioned in the introduction, these equations are well studied in the literature \cite{RishetikhinbqKZ,RishetikhinBqKZ2,Kastani}. By solving these equations, one obtains the explicit form of the amplitude $f^{\{R\},N...1}(x_1-t,...,x_j-t,...,x_N-t)$, using which the rest of the amplitudes in the wavefunction can be obtained by the action of the S-matrices discussed above. We shall present the explicit form of the wavefunction and the subsequent analysis uncovering dynamical phenomena associated with the boundaries in the forthcoming work.

\section{The case of the $U(1)$ Thirring model}
In the previous section, we have constructed the exact wavefunction that is the exact solution to the time-dependent Schrodinger equation of the $SU(2)$ Gross-Nevu model with integrable time-dependent bulk interaction strength, and derived the corresponding integrable time-dependent boundary conditions. We have shown that the consistency of the wavefunction gives rise to bqKZ equations corresponding to the rational XXX R-matrix. In this section, we shall discuss the case of the $U(1)$ Thirring model. We shall present the integrable boundary conditions and show that consistency of the wavefunction gives rise to the bqKZ equations corresponding to the XXZ R-matrix. 

Let us consider the general case of $N$ particles. The S-matrix corresponding to the interaction of a right mover and a left mover takes the same form as (\ref{2psmat}), 
but now with $R_{ij}(\lambda)$ being the XXZ R-matrix 
\begin{align}R_{ij}(\lambda)=\begin{pmatrix} 1&0&0&0\\0&\frac{\sinh(\lambda)}{\sinh(\lambda+iu)}&\frac{i\sin(u)}{\sinh(\lambda+iu)}&0\\0&\frac{i\sin(u)}{\sinh(\lambda+iu)}&\frac{\sinh(\lambda)}{\sinh(\lambda+iu)}&0\\0&0&0&1\end{pmatrix},\label{XXZRmat}\end{align}
where $u$ is the crossing parameter which is just the RG-invariant of the static model (\ref{utrel2}). As mentioned above, in the time-dependent model it corresponds to the dynamical invariant. The S-matrix corresponding to the exchange of the particles with the same chirality takes the same form as before (\ref{rel2}) and (\ref{rel3}), with $R_{ij}(\lambda)$ being the XXZ R-matrix (\ref{XXZRmat}).
These S-matrices satisfy the Yang-Baxter equations (\ref{YB1}) and (\ref{YB2}). Considering the interaction of the particles with the boundaries, one obtains the same reflection equations as before (\ref{reflectionL}) and (\ref{reflectionR}), but now with the bulk S-matrices taking the form of the XXZ R-matrix. The boundary conditions are related to the boundary K-matrices through the same relations as in (\ref{XXXbcl}) and (\ref{XXXbcr}), through which one can obtain the integrable boundary conditions once the boundary K-matrices satisfying the reflection equations are identified. The most general diagonal K-matrices of interest which satisfy the reflection equations corresponding to the case of XXZ R-matrix were obtained in \cite{RishetikhinbqKZ}, which provide the following integrable boundary conditions 
\begin{align}\mathcal{B}^{\dagger \chi}(t)=\begin{pmatrix} \sinh(i\xi^{\chi}+f(t))&0\\0&\sinh(i\xi^{\chi}-f(t))\end{pmatrix}, \chi=L,R.\label{XXZbc}\end{align}
Just as in the case of the SU(2) Gross-Neveu model, here the parameters $\xi^L,\xi^R$ are RG invariants in the static model and characterize the boundary phases exhibited by the system and their values lead to boundary eigenstate phase transitions \cite{PAA2}. In the time-dependent model they are dynamical invariants. Following the same procedure as in the case of the $SU(2)$ Gross-Neveu model, one can construct the transport operator which moves the particle through the entire system once. One obtains the bqKZ equations (\ref{BqKZ1}), (\ref{BqKZ2}) corresponding to the XXZ R-matrix with the bulk and the boundary S-matrices given by (\ref{XXZRmat}) and (\ref{XXZbc}).

In both the $U(1)$ Thirring model and the $SU(2)$ Gross-Neveu model, the bulk S-matrices are related to the corresponding R-matrices, whose argument is the function $f(t)$, which is associated with the spectral parameter.  Even though the relation between this function and the time-dependent interaction strengths is model dependent, in both cases it is linear in time (\ref{intrelf}). This is due to the fact that both the R-matrices corresponding to the $SU(2)$ Gross-Nevu model and the $U(1)$ Thirring model, which are the XXX and XXZ R-matrices respectively, are of the \textit{difference type}. Thus, it is natural to expect that the framework when applied to systems with non difference type R-matrices should give rise to more complicated functions for $f(t)$. Note that unlike the $SU(2)$ Gross-Neveu model, the $U(1)$ Thirring model exhibits an RG invariant-dynamical invariant in the bulk, which is the parameter $u$. Through bosonization, one obtains two decoupled sine-Gordon models: the charge degrees of freedom are described by a compact boson CFT whereas the spin degrees of freedom are described by the sine-Gordon model. The parameter $u$ in the $U(1)$ Thirring model corresponds to the radius of the compact boson $K$ of the sine-Gordon model and from the time dependent integrability-RG flow connection, one naturally expects that $K$ remains constant in the time-dependent sine-Gordon model. The time-dependent sine-Gordon model was studied in \cite{Benhoare}, where the generalized Lax connection that satisfies the zero curvature condition was constructed, using which it shown that the radius of the compact boson remains fixed while the coefficient of the cosine potential flows, which precisely agrees with the generalized Bethe ansatz results for the case of periodic boundary conditions discussed above. Thus extending the construction of Lax connection for the case of open boundary conditions following the method of \cite{Benhoare} should give rise to the integrable boundary conditions (\ref{XXZbc}) obtained through the generalized Bethe ansatz construction developed in this work.

\section{Discussion}
In this work we have considered the time-dependent $SU(2)$ Gross-Neveu model with boundary conditions that are also time-dependent. We have constructed the exact solution to the time-dependent Schrodinger equations and for integrable time-dependent bulk interaction strength, we derived constraints on the boundary conditions for the system to be integrable, which are imposed by the reflection equations. For boundary conditions satisfying these constraints, we have shown that the consistency of the wavefunction gives rise to a set of matrix difference equations which take the form of the Boundary quantum Knizhnik-Zamolodchikov equations. In the case of periodic boundary conditions, one instead obtains quantum Knizhnik-Zamolodchikov equations. Thus we have extended the generalized Bethe ansatz framework \cite{PasnooriKondo} to the case of open time-dependent boundary conditions. These time-dependent boundary conditions may be realized physically by applying time-varying magnetic fields close to the boundaries which induces a Zeeman effect \cite{PAA2}.

The method developed in this work is general and can be straightforwardly extended to identify and solve other time-dependent integrable models, as we briefly elaborate below. For a given integrable model, the algebra realized by the generators of the interactions determines the structure of the bulk S-matrix. The S-matrix entering the construction of the wavefunction \footnote{Here, we refer to the bare S-matrices entering the construction of the wavefunction, as opposed to the physical S-matrices governing the scattering of physical excitations.} can in turn be mapped to an R-matrix, which is generally a function of a spectral parameter taking values in the complex plane. In addition to the spectral parameter, the R-matrix contains a constant parameter, which is referred to as the crossing parameter. This parameter is generally associated with an invariant of the model: for time-independent systems it is related to an RG invariant, while for time-dependent systems it corresponds to a dynamical invariant \cite{PasnooriRG}. It is the parameter $u$ in the case of the $U(1)$ Thirring model discussed in this work. In contrast to the crossing parameter, the spectral parameter is a variable. For time-independent models, it is typically related to the interaction strength, whereas in the time-dependent case it becomes a function of the time-dependent interaction strength. It is the function $f(t)$ in the two cases discussed in this work. 

This observation provides a direct prescription for constructing the solution of the time-dependent model from the corresponding static model or equivalently, its time-independent counterpart. Given an integrable model with constant interaction strengths, one first identifies the corresponding R-matrix. The spectral parameter, which depends on the constant interaction strength, can then be promoted to a function of the time-dependent strength. For a given reference amplitude in the wavefunction, the time entering this function associated with the spectral parameter is chosen to be the time at which the corresponding particles interact, thereby fixing the S-matrix associated with their scattering. The same construction applies to scattering from a boundary. The boundary S-matrices are mapped to K-matrices, whose spectral parameters are similarly related to the interaction strengths, while their constant parameters, such as $\xi_{L,R}$ in the present work, are related to RG invariants, or equivalently to dynamical invariants in the time-dependent problem.

One can then transport a chosen particle through the system, moving it successively across the other particles and reflect off the boundaries. At each crossing, the above prescription determines the corresponding S-matrix, and hence the transport operator. The S-matrices not fixed directly by the Hamiltonian are constrained by the Yang-Baxter equation, which is satisfied by the bulk S-matrices. The bulk and boundary S-matrices are in turn constrained by the reflection equation. These consistency conditions ensure that the system is integrable and transporting a particle through the entire system then leads to a set of matrix difference equations involving the bulk and boundary S-matrices, or equivalently the corresponding R and K-matrices. Solving these difference equations yields the exact many-body wavefunction.

The above construction can in principle be applied to any integrable model with constant interaction strengths that is solvable by the regular Bethe ansatz and for which the corresponding R and K-matrices are known \cite{Malara_2006,Lima-Santos_2009,BATCHELOR,DeVega} and thus construct the corresponding time-dependent integrable models. Natural examples include higher-dimensional representations of $SU(2)$ as well as $SU(N)$ symmetric models. The case of $\mathcal{U}_q(\widehat{\text{sl}_2})$ is particularly interesting because it exhibits a rich variety of nontrivial phenomena in the case of constant strengths. For the spin-1 representation, one obtains the Zamolodchikov-Fateev 19-vertex model and the Izergin-Korepin model \cite{Lima-Santos_2009,LIMASANTOS}, while the spin-$1/2$ representation discussed in this work is also nontrivial as its corresponding static model exhibits symmetry-protected topological (SPT) phases \cite{PAA1,PAA2}.

We finally note that some care is required when applying the construction to models with quadratic dispersion or to systems defined on discrete spaces, such as lattice models. Nevertheless, the underlying principles remain applicable. We defer a detailed treatment of these cases to future work.

\bibliography{refpaper}

@article{cherednikbqkz,
	author = {Cherednik, Ivan},
	date = {1992/11/01},
	doi = {10.1007/BF02096568},
	id = {Cherednik1992},
	isbn = {1432-0916},
	journal = {Communications in Mathematical Physics},
	number = {1},
	pages = {109--136},
	title = {Quantum Knizhnik-Zamolodchikov equations and affine root systems},
	url = {https://doi.org/10.1007/BF02096568},
	volume = {150},
	year = {1992}}

@inproceedings{pasnoori2026sgcircuit,
  author    = {Pasnoori, Parameshwar R. and Azaria, Patrick and Sardashti, Kasra},
  title     = {Simulating sine-Gordon model with time dependent parameters using Josephson junction arrays},
  booktitle = {APS Global Physics Summit 2026},
  year      = {2026},
  month     = {March},
  note      = {Presented at the American Physical Society Global Physics Summit},
  url       = {https://meetings-archive.aps.org/smt/2026/mar-c18/10/}
}

@article{Cherednik1,
	author = {Cherednik, I.  V. },
	date = {1984/10/01},
	doi = {10.1007/BF01038545},
	id = {Cherednik1984},
	isbn = {1573-9333},
	journal = {Theoretical and Mathematical Physics},
	number = {1},
	pages = {977--983},
	title = {Factorizing particles on a half-line and root systems},
	url = {https://doi.org/10.1007/BF01038545},
	volume = {61},
	year = {1984}}

@article{BAJNOK2006179,
title = {On the boundary form factor program},
journal = {Nuclear Physics B},
volume = {750},
number = {3},
pages = {179-212},
year = {2006},
issn = {0550-3213},
doi = {https://doi.org/10.1016/j.nuclphysb.2006.05.019},
url = {https://www.sciencedirect.com/science/article/pii/S0550321306004342},
author = {Z. Bajnok and L. Palla and G. Takacs}}

@article{DiFrancesco_2007,
doi = {10.1088/1742-5468/2007/12/P12009},
url = {https://doi.org/10.1088/1742-5468/2007/12/P12009},
year = {2007},
month = {dec},
publisher = {IOP Publishing Ltd},
volume = {2007},
number = {12},
pages = {P12009},
author = {Di Francesco, P and Zinn-Justin, P},
title = {Quantum Knizhnik-Zamolodchikov equation: reflecting boundary conditions and combinatorics},
journal = {Journal of Statistical Mechanics: Theory and Experiment}
}

@article{Benhoare,
	author = {Hoare, Ben and Levine, Nat and Tseytlin, Arkady A.},
	date = {2020/11/06},
	doi = {10.1007/JHEP11(2020)020},
	id = {Hoare2020},
	isbn = {1029-8479},
	journal = {Journal of High Energy Physics},
	number = {11},
	pages = {20},
	title = {Sigma models with local couplings: a new integrability-RG flow connection},
	url = {https://doi.org/10.1007/JHEP11(2020)020},
	volume = {2020},
	year = {2020}}

@misc{pasnoorigrossneveu2,
      title={Time-Dependent Dynamical Dimensional Transmutation in the $SU(2)$ Gross-Neveu Model with Time-Dependent Interaction Strength}, 
      author={Parameshwar R. Pasnoori},
      year={2026},
      eprint={2605.05111},
      archivePrefix={arXiv},
      primaryClass={math-ph},
      url={https://arxiv.org/abs/2605.05111}, 
}

@article{PasnooriGrossNeveu,
  title = {Quantum Knizhnik-Zamolodchikov equations and integrability of quantum field theories with time-dependent interaction strength},
  author = {Pasnoori, Parameshwar R.},
  journal = {Phys. Rev. B},
  volume = {113},
  issue = {9},
  pages = {094514},
  numpages = {15},
  year = {2026},
  month = {Mar},
  publisher = {American Physical Society},
  doi = {10.1103/cr3y-71lr},
  url = {https://link.aps.org/doi/10.1103/cr3y-71lr}
}

@article{PasnooriKondo,
  title = {Integrability of the Kondo model with time-dependent interaction strength},
  author = {Pasnoori, Parameshwar R.},
  journal = {Phys. Rev. B},
  volume = {112},
  issue = {6},
  pages = {L060409},
  numpages = {6},
  year = {2025},
  month = {Aug},
  publisher = {American Physical Society},
  doi = {10.1103/78xb-5lmw},
  url = {https://link.aps.org/doi/10.1103/78xb-5lmw}
}

@article{PasnooriRG,
  title = {Quantum integrability of Hamiltonians with time-dependent interaction strengths and the renormalization group flow},
  author = {Pasnoori, Parameshwar R.},
  journal = {Phys. Rev. B},
  volume = {113},
  issue = {20},
  pages = {L201405},
  numpages = {6},
  year = {2026},
  month = {May},
  publisher = {American Physical Society},
  doi = {10.1103/yrmd-12wy},
  url = {https://link.aps.org/doi/10.1103/yrmd-12wy}
}

@article{RishetikhinbqKZ,
	author = {Reshetikhin, Nicolai and Stokman, Jasper and Vlaar, Bart},
	date = {2015/06/01},
	doi = {10.1007/s00220-014-2227-2},
	id = {Reshetikhin2015},
	isbn = {1432-0916},
	journal = {Communications in Mathematical Physics},
	number = {2},
	pages = {953--986},
	title = {Boundary Quantum Knizhnik--Zamolodchikov Equations and Bethe Vectors},
	url = {https://doi.org/10.1007/s00220-014-2227-2},
	volume = {336},
	year = {2015}}

@article{RishetikhinBqKZ2,
title = {Integral solutions to boundary quantum KnizhnikÃ¢ÂÂZamolodchikov equations},
journal = {Advances in Mathematics},
volume = {323},
pages = {486-528},
year = {2018},
issn = {0001-8708},
doi = {https://doi.org/10.1016/j.aim.2017.10.041},
url = {https://www.sciencedirect.com/science/article/pii/S0001870816302912},
author = {Nicolai Reshetikhin and Jasper Stokman and Bart Vlaar}
}

@inbook{Kastani,
author = {MASAHIRO KASATANI},
title = {BOUNDARY QUANTUM KNIZHNIK-ZAMOLODCHIKOV EQUATION},
booktitle = {New Trends in Quantum Integrable Systems},
chapter = {},
pages = {157-171},
doi = {10.1142/9789814324373_0009},
URL = {https://www.worldscientific.com/doi/abs/10.1142/9789814324373_0009}
}

@misc{komatsu2026,
      title={Time-Dependent Integrability from Gauge Theory, I}, 
      author={Shota Komatsu and Jun-ichi Sakamoto and Anders Wallberg and Masahito Yamazaki},
      year={2026},
      eprint={2607.02648},
      archivePrefix={arXiv},
      primaryClass={hep-th},
      url={https://arxiv.org/abs/2607.02648}, 
}

@article{Costello1,
    author = "Costello, Kevin and Witten, Edward and Yamazaki, Masahito",
    title = "{Gauge Theory and Integrability, I}",
    eprint = "1709.09993",
    archivePrefix = "arXiv",
    primaryClass = "hep-th",
    reportNumber = "IPMU17-0136",
    doi = "10.4310/ICCM.2018.v6.n1.a6",
    journal = "ICCM Not.",
    volume = "06",
    number = "1",
    pages = "46--119",
    year = "2018"
}

@article{Costello2,
    author = "Costello, Kevin and Witten, Edward and Yamazaki, Masahito",
    title = "{Gauge Theory and Integrability, II}",
    eprint = "1802.01579",
    archivePrefix = "arXiv",
    primaryClass = "hep-th",
    reportNumber = "IPMU18-0025",
    doi = "10.4310/ICCM.2018.v6.n1.a7",
    journal = "ICCM Not.",
    volume = "06",
    number = "1",
    pages = "120--146",
    year = "2018"
}

@misc{costello3,
      title={Gauge Theory And Integrability, III}, 
      author={Kevin Costello and Masahito Yamazaki},
      year={2019},
      eprint={1908.02289},
      archivePrefix={arXiv},
      primaryClass={hep-th},
      url={https://arxiv.org/abs/1908.02289}, 
}

@article{Malara_2006,
doi = {10.1088/1742-5468/2006/09/P09013},
url = {https://doi.org/10.1088/1742-5468/2006/09/P09013},
year = {2006},
month = {sep},
publisher = {},
volume = {2006},
number = {09},
pages = {P09013},
author = {Malara, R and Lima-Santos, A},
title = {On , , , , , , and  reflection K-matrices},
journal = {Journal of Statistical Mechanics: Theory and Experiment}
}

@article{PasnooriXXZPD,
  title = {Complete boundary phase diagram of the spin-$\frac{1}{2}$ XXZ chain with boundary fields in the antiferromagnetic gapped regime},
  author = {Pasnoori, Parameshwar R. and Tang, Yicheng and Lee, Junhyun and Pixley, J. H. and Azaria, Patrick and Andrei, Natan},
  journal = {Phys. Rev. B},
  volume = {113},
  issue = {5},
  pages = {054424},
  numpages = {31},
  year = {2026},
  month = {Feb},
  publisher = {American Physical Society},
  doi = {10.1103/5j68-1pqh},
  url = {https://link.aps.org/doi/10.1103/5j68-1pqh}
}

@article{XXZKondo,
  title = {Edge modes and boundary impurities in the anisotropic Heisenberg spin chain},
  author = {Kattel, Pradip and Pasnoori, Parameshwar R. and Pixley, J. H. and Andrei, Natan},
  journal = {Phys. Rev. B},
  volume = {111},
  issue = {17},
  pages = {174430},
  numpages = {36},
  year = {2025},
  month = {May},
  publisher = {American Physical Society},
  doi = {10.1103/PhysRevB.111.174430},
  url = {https://link.aps.org/doi/10.1103/PhysRevB.111.174430}
}

@article{AndreiLowenstein79,
  title = {Diagonalization of the Chiral-Invariant Gross-Neveu Hamiltonian},
  author = {Andrei, N. and Lowenstein, J. H.},
  journal = {Phys. Rev. Lett.},
  volume = {43},
  issue = {23},
  pages = {1698--1701},
  numpages = {0},
  year = {1979},
  month = {Dec},
  publisher = {American Physical Society},
  doi = {10.1103/PhysRevLett.43.1698},
  url = {https://link.aps.org/doi/10.1103/PhysRevLett.43.1698}
}

@article{Smirnov_1986,
doi = {10.1088/0305-4470/19/10/003},
url = {https://dx.doi.org/10.1088/0305-4470/19/10/003},
year = {1986},
month = {jul},
publisher = {IOP publishing},
volume = {19},
number = {10},
pages = {L575},
author = {F A Smirnov},
title = {A general formula for soliton form factors in the quantum sine-Gordon model},
journal = {Journal of Physics A: Mathematical and General}
}

@article{Tarasov:1993vs,
    author = "Tarasov, V. and Varchenko, A.",
    title = "{Jackson integral representations for solutions of the quantized Knizhnik-Zamolodchikov equation}",
    eprint = "hep-th/9311040",
    archivePrefix = "arXiv",
    reportNumber = "RIMS-949",
    month = "8",
    year = "1993"
}

@article{Tarasov:1994bb,
    author = "Tarasov, V. and Varchenko, A.",
    title = "{Asymptotic solutions to the quantized Knizhnik-Zamolodchikov equation and Bethe vectors}",
    eprint = "hep-th/9406060",
    archivePrefix = "arXiv",
    month = "5",
    year = "1994"
}

@article{PAA2,
  title = {Boundary-induced topological and mid-gap states in charge conserving one-dimensional superconductors: Fractionalization transition},
  author = {Pasnoori, Parameshwar R. and Andrei, Natan and Azaria, Patrick},
  journal = {Phys. Rev. B},
  volume = {104},
  issue = {13},
  pages = {134519},
  numpages = {17},
  year = {2021},
  month = {Oct},
  publisher = {American Physical Society},
  doi = {10.1103/PhysRevB.104.134519},
  url = {https://link.aps.org/doi/10.1103/PhysRevB.104.134519}
}

@article{PAA1,
  title = {Edge modes in one-dimensional topological charge conserving spin-triplet superconductors: Exact results from Bethe ansatz},
  author = {Pasnoori, Parameshawar R. and Andrei, Natan and Azaria, Patrick},
  journal = {Phys. Rev. B},
  volume = {102},
  issue = {21},
  pages = {214511},
  numpages = {17},
  year = {2020},
  month = {Dec},
  publisher = {American Physical Society},
  doi = {10.1103/PhysRevB.102.214511},
  url = {https://link.aps.org/doi/10.1103/PhysRevB.102.214511}
}

@article{WiegmannGN,
  title = {Peierls Transition in Gross-Neveu Model from Bethe Ansatz},
  author = {Melin, Valdemar and Sekiguchi, Yuta and Wiegmann, Paul and Zarembo, Konstantin},
  journal = {Phys. Rev. Lett.},
  volume = {133},
  issue = {10},
  pages = {101601},
  numpages = {6},
  year = {2024},
  month = {Sep},
  publisher = {American Physical Society},
  doi = {10.1103/PhysRevLett.133.101601},
  url = {https://link.aps.org/doi/10.1103/PhysRevLett.133.101601}
}

@article{TsvelickWiegmann1983,
  author       = {Tsvelick, A. M. and Wiegmann, P. B.},
  title        = {Exact results in the theory of magnetic alloys},
  journal      = {Advances in Physics},
  volume       = {32},
  number       = {4},
  pages        = {453--713},
  year         = {1983},
  publisher    = {Taylor & Francis},
  doi          = {10.1080/00018738300101561}
}

@phdthesis{Hulthen,
   author = {Hulthen, Lamek},
   institution = {Stockholm College},
   pages = {106},
   school = {, Stockholm College},
   title = {{\"U}ber das Austauschproblem eines Kristalles},
   series = {Arkiv fur matematik, astronomi och fysik},
   number = {26A:11},
   URL = {https://pubs.sub.su.se/1623.pdf},
   year = {1938}
}

@article{Frenkel,
	author = {Frenkel, I. B. and Reshetikhin, N. Yu.},
	date = {1992/05/01},
	doi = {10.1007/BF02099206},
	id = {Frenkel1992},
	isbn = {1432-0916},
	journal = {Communications in Mathematical Physics},
	number = {1},
	pages = {1-60},
	title = {Quantum affine algebras and holonomic difference equations},
	url = {https://doi.org/10.1007/BF02099206},
	volume = {146},
	year = {1992}}

@misc{pasnoori2025interplay,
      title={Interplay between Symmetry Breaking and Interactions in a Symmetry Protected Topological Phase}, 
      author={Parameshwar R. Pasnoori and Patrick Azaria},
      year={2025},
      eprint={2506.19771},
      archivePrefix={arXiv},
      primaryClass={hep-th},
      url={https://arxiv.org/abs/2506.19771}, 
}

@misc{pasnoori2025duality,
      title={Duality symmetry, zero energy modes and boundary spectrum of the sine-Gordon/massive Thirring model}, 
      author={Parameshwar R. Pasnoori and Ari Mizel and Patrick Azaria},
      year={2025},
      eprint={2503.14776},
      archivePrefix={arXiv},
      primaryClass={hep-th},
      url={https://arxiv.org/abs/2503.14776}, 
}

@article{Keselman,
  title = {Gapless symmetry-protected topological phase of fermions in one dimension},
  author = {Keselman, Anna and Berg, Erez},
  journal = {Phys. Rev. B},
  volume = {91},
  issue = {23},
  pages = {235309},
  numpages = {12},
  year = {2015},
  month = {Jun},
  publisher = {American Physical Society},
  doi = {10.1103/PhysRevB.91.235309},
  url = {https://link.aps.org/doi/10.1103/PhysRevB.91.235309}
}

@article{rishetikhin1,
	author = {Reshetikhin, N. },
	date = {1992/11/01},
	doi = {10.1007/BF00420749},
	id = {Reshetikhin1992},
	isbn = {1573-0530},
	journal = {Letters in Mathematical Physics},
	number = {3},
	pages = {153-165},
	title = {Jackson-type integrals, Bethe vectors, and solutions to a difference analog of the Knizhnik-Zamolodchikov system},
	url = {https://doi.org/10.1007/BF00420749},
	volume = {26},
	year = {1992}}

@article{rishetikhin2,
	author = {Reshetikhin, Nicolai},
	date = {1992/11/01},
	doi = {10.1007/BF00420750},
	id = {Reshetikhin1992},
	isbn = {1573-0530},
	journal = {Letters in Mathematical Physics},
	number = {3},
	pages = {167-177},
	title = {The Knizhnik-Zamolodchikov system as a deformation of the isomonodromy problem},
	url = {https://doi.org/10.1007/BF00420750},
	volume = {26},
	year = {1992}}

@misc{pasnoori2025Kondo2,
      title={Exact many-body wavefunction of the Kondo model with time-dependent interaction strength}, 
      author={Parameshwar R. Pasnoori and Emil. A. Yuzbashyan},
      year={2025},
      eprint={2509.05640},
      archivePrefix={arXiv},
      primaryClass={cond-mat.str-el},
      url={https://arxiv.org/abs/2509.05640}, 
}

@article{Thouless,
  title = {Quantized Hall Conductance in a Two-Dimensional Periodic Potential},
  author = {Thouless, D. J. and Kohmoto, M. and Nightingale, M. P. and den Nijs, M.},
  journal = {Phys. Rev. Lett.},
  volume = {49},
  issue = {6},
  pages = {405--408},
  numpages = {0},
  year = {1982},
  month = {Aug},
  publisher = {American Physical Society},
  doi = {10.1103/PhysRevLett.49.405},
  url = {https://link.aps.org/doi/10.1103/PhysRevLett.49.405}
}

@article{Laughlin,
  title = {Anomalous Quantum Hall Effect: An Incompressible Quantum Fluid with Fractionally Charged Excitations},
  author = {Laughlin, R. B.},
  journal = {Phys. Rev. Lett.},
  volume = {50},
  issue = {18},
  pages = {1395--1398},
  numpages = {0},
  year = {1983},
  month = {May},
  publisher = {American Physical Society},
  doi = {10.1103/PhysRevLett.50.1395},
  url = {https://link.aps.org/doi/10.1103/PhysRevLett.50.1395}
}

@article{Yasuhiro,
  title = {Chern number and edge states in the integer quantum Hall effect},
  author = {Hatsugai, Yasuhiro},
  journal = {Phys. Rev. Lett.},
  volume = {71},
  issue = {22},
  pages = {3697--3700},
  numpages = {0},
  year = {1993},
  month = {Nov},
  publisher = {American Physical Society},
  doi = {10.1103/PhysRevLett.71.3697},
  url = {https://link.aps.org/doi/10.1103/PhysRevLett.71.3697}
}

@article{halperin,
  title = {Quantized Hall conductance, current-carrying edge states, and the existence of extended states in a two-dimensional disordered potential},
  author = {Halperin, B. I.},
  journal = {Phys. Rev. B},
  volume = {25},
  issue = {4},
  pages = {2185--2190},
  numpages = {0},
  year = {1982},
  month = {Feb},
  publisher = {American Physical Society},
  doi = {10.1103/PhysRevB.25.2185},
  url = {https://link.aps.org/doi/10.1103/PhysRevB.25.2185}
}

@article{kellendonk,
author = {KELLENDONK, J. and RICHTER, T. and SCHULZ-BALDES, H.},
title = {EDGE CURRENT CHANNELS AND CHERN NUMBERS IN THE INTEGER QUANTUM HALL EFFECT},
journal = {Reviews in Mathematical Physics},
volume = {14},
number = {01},
pages = {87-119},
year = {2002},
doi = {10.1142/S0129055X02001107},
URL = { https://doi.org/10.1142/S0129055X02001107}
}

@article{Pollman,
  title = {Detection of symmetry-protected topological phases in one dimension},
  author = {Pollmann, Frank and Turner, Ari M.},
  journal = {Phys. Rev. B},
  volume = {86},
  issue = {12},
  pages = {125441},
  numpages = {13},
  year = {2012},
  month = {Sep},
  publisher = {American Physical Society},
  doi = {10.1103/PhysRevB.86.125441},
  url = {https://link.aps.org/doi/10.1103/PhysRevB.86.125441}
}

@article{LIMASANTOS,
title = {Reflection K-matrices for 19-vertex models},
journal = {Nuclear Physics B},
volume = {558},
number = {3},
pages = {637-667},
year = {1999},
issn = {0550-3213},
doi = {https://doi.org/10.1016/S0550-3213(99)00456-3},
url = {https://www.sciencedirect.com/science/article/pii/S0550321399004563},
author = {A. Lima-Santos}
}

@article{deVega,
doi = {10.1088/0305-4470/26/12/007},
url = {https://doi.org/10.1088/0305-4470/26/12/007},
year = {1993},
month = {jun},
publisher = {},
volume = {26},
number = {12},
pages = {L519},
author = {H J de Vega and A Gonzalez Ruiz},
title = {Boundary K-matrices for the six vertex and the n(2n-1)An-1 vertex models},
journal = {Journal of Physics A: Mathematical and General}
}

@article{INAMI,
title = {Reflection K-matrices of the 19-vertex model and XXZ spin-1 chain with general boundary terms},
journal = {Nuclear Physics B},
volume = {470},
number = {3},
pages = {419-432},
year = {1996},
issn = {0550-3213},
doi = {https://doi.org/10.1016/0550-3213(96)00133-2},
url = {https://www.sciencedirect.com/science/article/pii/0550321396001332},
author = {Takeo Inami and Satoru Odake and Yao-Zhong Zhang}
}

@article{BATCHELOR,
title = {Solutions of the reflection equation for face and vertex models associated with An(1), Bn(1), Cn(1), Dn(1) and An(2)},
journal = {Physics Letters B},
volume = {376},
number = {4},
pages = {266-274},
year = {1996},
issn = {0370-2693},
doi = {https://doi.org/10.1016/0370-2693(96)00319-X},
url = {https://www.sciencedirect.com/science/article/pii/037026939600319X},
author = {M.T Batchelor and V Fridkin and A Kuniba and Y.K Zhou}
}

@article{Lima-Santos_2009,
doi = {10.1088/1742-5468/2009/07/P07045},
url = {https://doi.org/10.1088/1742-5468/2009/07/P07045},
year = {2009},
month = {jul},
publisher = {},
volume = {2009},
number = {07},
pages = {P07045},
author = {Lima-Santos, A},
title = {Reflection matrices for the Uq[osp(r|2m)(1)] 
vertex model},
journal = {Journal of Statistical Mechanics: Theory and Experiment}
}

@misc{pasnooriNHKtime,
      title={Beyond Integrability Preserving Renormalization-Group Protocol in Non-Hermitian Hamiltonians with Time-Dependent Interaction Strengths}, 
      author={Parameshwar R. Pasnoori},
      year={2026},
      eprint={2608.19519},
      archivePrefix={arXiv},
      primaryClass={quant-ph},
      url={https://arxiv.org/abs/2608.19519}, 
}

@article{EVtrotter,
  title = {Integrable Digital Quantum Simulation: Generalized Gibbs Ensembles and Trotter Transitions},
  author = {Vernier, Eric and Bertini, Bruno and Giudici, Giuliano and Piroli, Lorenzo},
  journal = {Phys. Rev. Lett.},
  volume = {130},
  issue = {26},
  pages = {260401},
  numpages = {7},
  year = {2023},
  month = {Jun},
  publisher = {American Physical Society},
  doi = {10.1103/PhysRevLett.130.260401},
  url = {https://link.aps.org/doi/10.1103/PhysRevLett.130.260401}
}

@article{EVcircuitSM,
  title = {Strong Zero Modes in Integrable Quantum Circuits},
  author = {Vernier, Eric and Yeh, Hsiu-Chung and Piroli, Lorenzo and Mitra, Aditi},
  journal = {Phys. Rev. Lett.},
  volume = {133},
  issue = {5},
  pages = {050606},
  numpages = {7},
  year = {2024},
  month = {Aug},
  publisher = {American Physical Society},
  doi = {10.1103/PhysRevLett.133.050606},
  url = {https://link.aps.org/doi/10.1103/PhysRevLett.133.050606}
}

@ARTICLE{Sklyanin,
   author = {{Sklyanin}, E.~K.},
    title = "{Boundary conditions for integrable quantum systems}",
  journal = {Journal of Physics A Mathematical General},
     year = {1988},
    month = {may},
   volume = {21},
    pages = {2375-2389},
      doi = {10.1088/0305-4470/21/10/015},
   adsurl = {http://adsabs.harvard.edu/abs/1988JPhA...21.2375S}
}

@article{NHbbcor,
	author = {Xiao, Lei and Deng, Tianshu and Wang, Kunkun and Zhu, Gaoyan and Wang, Zhong and Yi, Wei and Xue, Peng},
	date = {2020/07/01},
	doi = {10.1038/s41567-020-0836-6},
	id = {Xiao2020},
	isbn = {1745-2481},
	journal = {Nature Physics},
	number = {7},
	pages = {761--766},
	title = {Non-Hermitian bulk--boundary correspondence in quantum dynamics},
	url = {https://doi.org/10.1038/s41567-020-0836-6},
	volume = {16},
	year = {2020}}

@article{Kane,
  title = {Quantum Spin Hall Effect in Graphene},
  author = {Kane, C. L. and Mele, E. J.},
  journal = {Phys. Rev. Lett.},
  volume = {95},
  issue = {22},
  pages = {226801},
  numpages = {4},
  year = {2005},
  month = {Nov},
  publisher = {American Physical Society},
  doi = {10.1103/PhysRevLett.95.226801},
  url = {https://link.aps.org/doi/10.1103/PhysRevLett.95.226801}
}

@article{Haldane,
  title = {Model for a Quantum Hall Effect without Landau Levels: Condensed-Matter Realization of the "Parity Anomaly"},
  author = {Haldane, F. D. M.},
  journal = {Phys. Rev. Lett.},
  volume = {61},
  issue = {18},
  pages = {2015--2018},
  numpages = {0},
  year = {1988},
  month = {Oct},
  publisher = {American Physical Society},
  doi = {10.1103/PhysRevLett.61.2015},
  url = {https://link.aps.org/doi/10.1103/PhysRevLett.61.2015}
}

@article{BKT,
doi = {10.1088/0022-3719/6/7/010},
url = {https://doi.org/10.1088/0022-3719/6/7/010},
year = {1973},
month = {apr},
publisher = {},
volume = {6},
number = {7},
pages = {1181},
author = {J M Kosterlitz and D J Thouless},
title = {Ordering, metastability and phase transitions in two-dimensional systems},
journal = {Journal of Physics C: Solid State Physics}
}

@article{Babujian_1997,
doi = {10.1088/0305-4470/30/18/019},
url = {https://dx.doi.org/10.1088/0305-4470/30/18/019},
year = {1997},
month = {sep},
publisher = {IOP Publishing},
volume = {30},
number = {18},
pages = {6425},
author = {H Babujian and M Karowski and A Zapletal},
title = {Matrix difference equations and a nested Bethe ansatz},
journal = {Journal of Physics A: Mathematical and General}
}

@article{KnizhnikZamolodchikov,
title = {Current algebra and Wess-Zumino model in two dimensions},
journal = {Nuclear Physics B},
volume = {247},
number = {1},
pages = {83-103},
year = {1984},
issn = {0550-3213},
doi = {https://doi.org/10.1016/0550-3213(84)90374-2},
url = {https://www.sciencedirect.com/science/article/pii/0550321384903742},
author = {V.G. Knizhnik and A.B. Zamolodchikov}
}

@article{varchenko,
	author = {Varchenko, A.  N. },
	date = {1995/07/01},
	doi = {10.1007/BF02103772},
	id = {Varchenko1995},
	isbn = {1432-0916},
	journal = {Communications in Mathematical Physics},
	number = {1},
	pages = {99-137},
	title = {Asymptotic solutions to the Knizhnik-Zamolodchikov equation and crystal base},
	url = {https://doi.org/10.1007/BF02103772},
	volume = {171},
	year = {1995}}

@article{Bethe1931,
    author = {Bethe, H.},
    title = {Zur Theorie der Metalle},
    journal = {Zeitschrift fur Physik},
    volume = {71},
    pages = {205-226},
    year = {1931},
    doi={10.1007/BF01341708},
    url={https://doi.org/10.1007/BF01341708}
}

@article{Andrei80,
  title = {Diagonalization of the Kondo Hamiltonian},
  author = {Andrei, N.},
  journal = {Phys. Rev. Lett.},
  volume = {45},
  issue = {5},
  pages = {379-382},
  numpages = {0},
  year = {1980},
  month = {Aug},
  publisher = {American Physical Society},
  doi = {10.1103/PhysRevLett.45.379},
  url = {https://link.aps.org/doi/10.1103/PhysRevLett.45.379}
}

@article{Wiegmann_1981,
doi = {10.1088/0022-3719/14/10/014},
url = {https://dx.doi.org/10.1088/0022-3719/14/10/014},
year = {1981},
month = {apr},
publisher = {IOP Publishing},
volume = {14},
number = {10},
pages = {1463},
author = {P B Wiegmann},
title = {Exact solution of the s-d exchange model (Kondo problem)},
journal = {Journal of Physics C: Solid State Physics}
}

@article{Thacker,
  title = {Structure and solution of the massive Thirring model},
  author = {Bergknoff, H. and Thacker, H. B.},
  journal = {Phys. Rev. D},
  volume = {19},
  issue = {12},
  pages = {3666-3681},
  numpages = {0},
  year = {1979},
  month = {Jun},
  publisher = {American Physical Society},
  doi = {10.1103/PhysRevD.19.3666},
  url = {https://link.aps.org/doi/10.1103/PhysRevD.19.3666}
}

@article{SklyaninQISM,
	address = {USSR},
	author = {Sklyanin, E K and Takhtadzhyan, L. A. and Faddeev, L. D.},
	isbn = {0564-6162},
	journal = {Teoreticheskaya i Matematicheskaya Fizika},
	number = {2},
	pages = {194-220},
	series = {Kvantovyj metod obratnoj zadachi 1},
	title = {Quantum inverse problem method 1},
	url = {http://inis.iaea.org/search/search.aspx?orig_q=RN:11506395},
	volume = {40},
	year = {1979}}

@article{parmeshkondo1,
  title = {Kondo impurity at the edge of a superconducting wire},
  author = {Pasnoori, Parameshwar R. and Rylands, Colin and Andrei, Natan},
  journal = {Phys. Rev. Res.},
  volume = {2},
  issue = {1},
  pages = {013006},
  numpages = {14},
  year = {2020},
  month = {Jan},
  publisher = {American Physical Society},
  doi = {10.1103/PhysRevResearch.2.013006},
  url = {https://link.aps.org/doi/10.1103/PhysRevResearch.2.013006}
}

@article{parmeshkondo2,
  title = {Rise and fall of Yu-Shiba-Rusinov bound states in charge-conserving s-wave one-dimensional superconductors},
  author = {Pasnoori, Parameshwar R. and Andrei, Natan and Rylands, Colin and Azaria, Patrick},
  journal = {Phys. Rev. B},
  volume = {105},
  issue = {17},
  pages = {174517},
  numpages = {13},
  year = {2022},
  month = {May},
  publisher = {American Physical Society},
  doi = {10.1103/PhysRevB.105.174517},
  url = {https://link.aps.org/doi/10.1103/PhysRevB.105.174517}
}

@article{KondoXXX,
  title = {Kondo effect in the isotropic Heisenberg spin chain},
  author = {Kattel, Pradip and Pasnoori, Parameshwar R. and Pixley, J. H. and Azaria, Patrick and Andrei, Natan},
  journal = {Phys. Rev. B},
  volume = {109},
  issue = {17},
  pages = {174416},
  numpages = {18},
  year = {2024},
  month = {May},
  publisher = {American Physical Society},
  doi = {10.1103/PhysRevB.109.174416},
  url = {https://link.aps.org/doi/10.1103/PhysRevB.109.174416}
}

@article{SUSYpaper,
  title = {Emergent boundary supersymmetry in a one-dimensional superconductor},
  author = {Pasnoori, Parameshwar R. and Azaria, Patrick and Rylands, Colin and Andrei, Natan},
  journal = {Phys. Rev. B},
  volume = {113},
  issue = {5},
  pages = {054509},
  numpages = {20},
  year = {2026},
  month = {Feb},
  publisher = {American Physical Society},
  doi = {10.1103/pzr7-cw4c},
  url = {https://link.aps.org/doi/10.1103/pzr7-cw4c}
}

@article{XXXpaper,
  title = {Boundary quantum phase transitions in the spin-$\frac{1}{2}$ Heisenberg chain with boundary magnetic fields},
  author = {Pasnoori, Parameshwar R. and Lee, Junhyun and Pixley, J. H. and Andrei, Natan and Azaria, Patrick},
  journal = {Phys. Rev. B},
  volume = {107},
  issue = {22},
  pages = {224412},
  numpages = {24},
  year = {2023},
  month = {Jun},
  publisher = {American Physical Society},
  doi = {10.1103/PhysRevB.107.224412},
  url = {https://link.aps.org/doi/10.1103/PhysRevB.107.224412}
}

@article{XXZpaper,
  title = {Spin fractionalization and zero modes in the spin-$\frac{1}{2}$ XXZ chain with boundary fields},
  author = {Pasnoori, Parameshwar R. and Tang, Yicheng and Lee, Junhyun and Pixley, J. H. and Andrei, Natan and Azaria, Patrick},
  journal = {Phys. Rev. B},
  volume = {112},
  issue = {7},
  pages = {075121},
  numpages = {6},
  year = {2025},
  month = {Aug},
  publisher = {American Physical Society},
  doi = {10.1103/thlq-h58t},
  url = {https://link.aps.org/doi/10.1103/thlq-h58t}
}

@misc{circuitSPT,
      title={Realizing a Symmetry Protected Topological Phase in a Superconducting Circuit}, 
      author={Parameshwar R. Pasnoori and Patrick Azaria and Ari Mizel},
      year={2025},
      eprint={2503.13406},
      archivePrefix={arXiv},
      primaryClass={quant-ph},
      url={https://arxiv.org/abs/2503.13406}, 
}

\end{document}